\documentclass[draft]{agujournal2019}
\usepackage{url} 
\usepackage{lineno}
\usepackage[inline]{trackchanges} 
\usepackage{soul}
\usepackage{float}
\usepackage{epstopdf, epsfig}
\usepackage{caption}
\usepackage{subcaption}
\usepackage{amsmath}
\usepackage{amssymb}
\usepackage{color}
\usepackage{placeins}
\usepackage{dcolumn}%
\usepackage{bm}%

\usepackage{enumitem}

\draftfalse

\journalname{Journal of Advances in Modeling Earth Systems (JAMES)}

\begin{document}

 \title{A non-intrusive machine learning framework for debiasing long-time coarse resolution climate simulations and quantifying rare events statistics }

%
%
\authors{B. Barthel Sorensen\affil{1}, A. Charalampopoulos\affil{1}, S. Zhang\affil{2}, B. E. Harrop\affil{2}, L. R. Leung\affil{2}, and T. P. Sapsis\affil{1}}

\affiliation{1}{Department of Mechanical Engineering, Massachusetts Institute of Technology, Cambridge, MA 02139, USA}
\affiliation{2}{Pacific Northwest National Laboratory, Richland, WA 99354, USA}

\correspondingauthor{T. P. Sapsis}{sapsis@mit.edu}



\begin{keypoints}
\item Development of non-intrusive correction operators for coarse scale climate simulations
\item Design of a training procedure that improves dynamics and allows for characterization of  extremes with return period longer than the training data
\item Application to Energy Exascale Earth System Model and demonstration of improvement on global and regional statistics
\end{keypoints}

\begin{abstract}
Due to the rapidly changing climate, the frequency and severity of extreme weather is expected to increase over the coming decades. As fully-resolved climate simulations remain computationally intractable, policy makers must rely on coarse-models to quantify risk for extremes. However, coarse models suffer from inherent bias due to the ignored “sub-grid” scales. We propose a framework to \textit{non-intrusively} debias coarse-resolution climate predictions using neural-network (NN) correction operators. Previous efforts have attempted to train such operators using loss functions that match statistics. However, this approach falls short with events that have longer return period than that of the training data, since the reference statistics have not converged. Here, the scope is to formulate a learning method that allows for correction of dynamics and quantification of extreme events with longer return period than the training data. The key obstacle is the chaotic nature of the underlying dynamics. To overcome this challenge, we introduce a dynamical systems approach where the correction operator is trained using reference data and a coarse model simulation \textit{nudged} towards that reference. The method is demonstrated on debiasing an under-resolved quasi-geostrophic model and the Energy Exascale Earth System Model (E3SM). For the former, our method enables the quantification of events that have \textcolor{black}{return period two orders longer} than the training data. For the latter, when trained on 8 years of ERA5 data, our approach is able to correct the coarse E3SM output to closely reflect the 36-year ERA5 statistics for all prognostic variables and significantly reduce their spatial biases. 
\end{abstract}

\section*{Plain Language Summary}
We present a general framework to design machine learned correction operators to improve the predicted statistics of low-resolution climate simulations. We illustrate the approach, which acts on existing data in a post-processing manner, on a simplified prototype climate model as well as a realistic climate model, namely the Energy Exascale Earth System Model (E3SM) with 110km resolution.  For the latter, we show that the developed approach is able to correct the low-resolution E3SM output to closely reflect the climate statistics \textcolor{black}{of historical observations as quantified by the ERA5 data set}.  We also demonstrate that our model significantly improves the prediction of atmospheric rivers, an example of extreme weather events resolvable by the low resolution model.

\section{Introduction}\label{sec:intro}
As climate changes, several studies have indicated that the frequency and severity of extreme weather events will increase over the coming decades \cite{raymond_understanding_2020,robinson_increasing_2021,fischer_increasing_2021}. Accurately quantifying the risk of such events is a critical step in developing strategies to prepare for and mitigate their negative impacts on society – which can include billions of dollars in damages and thousands of lost lives \cite{allen_managing_2012,houser_economic_2015,fiedler_business_2021}. However, predicting the risk, magnitude, and impacts of such events is difficult and multifaceted. First, these events are seldom observed and arise due to a range of -- often not fully understood -- physical mechanisms \cite{lucarini_extremes_2016,sapsis_statistics_2021}. Moreover, the most devastating events are those which arise due to extreme excursions of multiple variables simultaneously, such as concurrent drought and heatwaves, which have a combined effect greater than each would have had in isolation \cite{bevacqua_advancing_2023,zscheischler_future_2018,raymond_understanding_2020,robinson_increasing_2021}. In addition, these extremes, whether occurring in isolation or in concert, interact with the earth system -- and society -- in myriad and often non-trivial ways. For example, the aforementioned combination of excess heat and below-average precipitation can increase the frequency of wildfires, degrade soil quality, and intensify water shortages, all of which then in turn have devastating socioeconomic impacts through, for example, reduced crop yields and even increased spread of disease  \cite{barriopedro_hot_2011,witte_nasa_2011,hauser_role_2016,geirinhas_recent_2021}. Fully quantifying this complicated and interconnected system of physical, ecological, and social factors will surely require innovation and collaboration on a vast scale \cite{bauer_digital_2021,slingo_ambitious_2022}, yet even the first step, the accurate modeling of the climate dynamics, remains a challenging and unsolved problem.  

At their heart, climate models \cite{smagorinsky_general_1963,smagorinsky_numerical_1965,manabe_simulated_1965,mintz_very_1968}, or their more modern counterpart, Earth System Models (ESM) \cite{taylor2009non,dennis2012cam, golaz2022doe} are discretized forms of the equations of motion governing the Earth atmosphere and oceans. These known dynamical equations are then coupled to theoretical or empirical \textcolor{black}{parameterizations} of phenomena whose governing equations are unknown, such as the exact relationship between the vertical distribution of water vapor and precipitation rates \cite{stensrud_parameterization_2007,holloway_moisture_2009} or the residence time of carbon in various terrestrial reservoirs \cite{friend_carbon_2014,bloom_decadal_2016}. Statistical climate predictions are then made by averaging over ensembles of realizations generated by such models. Unfortunately, a significant challenge in the practical application of these models is the computational complexity incurred by the vast range of dynamically active scales present in the oceans and atmosphere. This challenge is compounded when considering the need for large ensembles of models to be run over time horizons stretching decades or even centuries. The current \textcolor{black}{state-of-the-art} for climate modeling corresponds  to an atmospheric spatial resolution of approximately 1 degree (i.e. approximately $110$ km), with some early progress seen in the development of $<5$ km resolution models \cite{tomita_global_2005,stevens_dyamond_2019,wedi_baseline_2020}. While there are some proponents of even finer ($1$ km) resolution simulations \cite{bauer_digital_2021,slingo_ambitious_2022}, even these fail to resolve critical phenomena such as the dynamics of stratocumulus clouds, which evolve on length scales of around 10 m \cite{wood_stratocumulus_2012,schneider_climate_2017}, much less than the Kolmogorov dissipation scale which is on the order of 1 mm. In fact, the degrees of freedom in an ESM with $1$ km resolution, which is stretching today's computational capabilities, fall short of what is needed to fully resolve atmospheric turbulence by a factor of $10^{17}$ \cite{schneider_harnessing_2023}. These realities imply that the brute-force computation of the climate system will remain out of reach for the foreseeable future and that meaningful progress will require new and innovative solutions.

One promising and growing area of research to sidestep the computational intractability of fully resolved simulations is the combination of existing climate models with modern machine learning (ML) and data-assimilation strategies which learn the ``sub-grid'' dynamics from targeted high resolution simulations or observational data \cite{schneider_earth_2017,schneider_harnessing_2023}. For example, reservoir-computing-based hybrid models have recently been demonstrated which learn online corrections to coarse climate models. These have been shown to substantially reduce overall bias \cite{arcomano_hybrid_2022} and capture events, such as sudden stratospheric warming, which are not resolved at all in free-running coarse climate models \cite{arcomano_hybrid_2023}. 
Another, and perhaps more widely adopted approach is the data-driven parametric closure model. Here ``closure model'' refers to a state-dependent forcing term which aims to mimic the dynamic effects of the un-resolved scales on the resolved ones. Initially, such strategies were demonstrated on idealized aqua planet configurations using random forests (RF) \cite{yuval_stable_2020} and neural network (NN) models \cite{rasp_deep_2018,brenowitz_spatially_2019,yuval_use_2021}. More recently they have been applied to realistic global climate models to learn parametric forcing terms from reanalysis data using RFs \cite{watt-meyer_correcting_2021} \textcolor{black}{and Deep Operator Networks (DeepONet) \cite{bora_learning_2023}, as well as from higher resolution} simulations with 3 km \cite{bretherton_correcting_2022}, and 25 km \cite{clark_correcting_2022} resolution -- both utilizing NNs and RFs. Across these studies, the ML closure models led to a robust improvement of $20-30\%$ in certain predicted integral quantities such as mean precipitation. However, predictions of other quantities were less reliable. For example, \cite{clark_correcting_2022} found that surface temperature predictions depended non-trivially on the random seed used in training the ML model. Furthermore, these approaches did not universally reduce the bias of the predicted climate over the uncorrected baseline, even in some cases increasing the bias of the coarse model \cite{watt-meyer_correcting_2021,clark_correcting_2022}. 

Despite these concerns, the most severe limitation of these approaches is numerical instability when integrating over long time horizons. This means that the aforementioned studies have only been demonstrated over short, 1 year \cite{watt-meyer_correcting_2021} and 5.25 year \cite{clark_correcting_2022} time horizons -- far shorter than what is required for long-term climate analysis. Such instabilities are inherent in this type of intrusive approach,  \textcolor{black}{except of special classes of representations for the closure terms which can guarantee stability of one-point and two-point statistics \cite{zhang_error_2021}}. The ML correction term augmenting the coarse-scale equations is designed to bring the turbulent attractor of the corrected system in line with that of the reference. However, the ML approximation of the sub-grid scale dynamics will not be perfect, and due to the chaotic nature of the system, small excursions will eventually grow, causing the predicted system trajectory to diverge from the attractor of the reference data \cite{wikner_stabilizing_2022}.  We refer the interested reader to \citeA{yuval_use_2021} for a detailed discussion of the stability challenges inherent in data-driven closure models.

Motivated by the intrinsic limitation of data-driven closure-models, we consider a different strategy. We seek to learn a ML operator which does not alter the equations, but rather acts as a post-processing operation to debias coarse scaled climate models. Such a \textit{non-intrusive} approach has several theoretical advantages. First, it does not require altering the code of the core climate model -- a non-trivial endeavour which often requires the harmonization of codes written in different languages \cite{mcgibbon_fv3gfs-wrapper_2021}. Second, unlike a closure model, it is domain agnostic, it can be applied globally or only for specific regions or altitudes. Third, and most critically, it is not susceptible to the same instabilities which plague schemes which apply machine learning corrections directly to the system dynamics. This in turn means it can be used to generate ensembles of trajectories over century $+$ time horizons -- a necessary step for quantifying risk of rare climate events with very long return periods. However, machine learning such a non-intrusive correction presents several considerable challenges, the foremost of which is the chaotic character of the climate systems under investigation. A mapping learned directly from some particular trajectory of a coarse model to a reference is unlikely to generalize, as it will encode not only the differences inherent in the coarse-scaling but it will also be corrupted by the particular chaotic realization of the training data. To overcome this challenge, \citeA{arbabi_generative_2022} developed a generative framework which uses a system of linear stochastic differential equations in conjunction with a nonlinear map modeled through optimal transport. The nonlinear map and the stochastic linear system are optimized so that the statistics of the output match the statistics of the training data. In a more recent work, \citeA{blanchard_multi-scale_2022} used a more complex architecture consisting of a spatial wavelet decomposition, a temporal-convolutional-network (TCN) and long-short-term-memory (LSTM) architectures trained also on a purely statistical loss function involving single point probability densities and temporal spectrum. \textcolor{black}{Alternatively, strategies such as generative adversarial networks (GAN) \cite{mcgibbon_global_2023} and unsupervised image-to-image networks (UNIT) \cite{fulton_bias_2023} have been used to correct biases in average precipitation rates -- an integral quantity which is less affected by stochastic variation.} While machine learning correction operators using a purely statistical loss function can indeed generate trajectories with plausible statistics, this property alone does not guarantee the resulted spatio-temporal dynamics are always physically realistic. Most importantly the quality of the resulted models, by design, cannot exceed the quality of the statistics used for training. Therefore, if the statistics for rare events of a given (large) return period have not converged (because of low availability of such events in the training set) the model is essentially forced to reproduce inaccurate, i.e. non-converged statistics, at least for rare events that have return periods comparable or longer than the training data set. To this end, methods based on purely  statistical loss functions cannot be used for statistical extrapolation.

In this work we describe a framework to overcome this challenge. Our aim it to design an algorithm that learns essential dynamics and is able to extrapolate statistics with a non-intrusive approach. The heart of the proposed strategy is that we do not machine learn a map from any \textit{arbitrary} coarse trajectory to the reference, but specifically from a coarse trajectory \textit{nudged towards that reference}. Nudging the coarse model towards the target reference trajectory results in an input trajectory which predominately obeys the dynamics of the coarse model yet remains close to the reference trajectory. Training a ML operator on this specific pair of trajectories allows us to learn a transformation which encodes only the differences caused by the coarse-grid without being corrupted by random stochastic effects. Once trained, this correction operator can then reliably map \textit{any} free-running coarse trajectory into the attractor of the reference data. We first lay out the theoretical framework of the proposed strategy in terms of a general chaotic dynamical system. We then illustrate our method on a simplified 2-layer quasi-geostrophic (QG) model, and show that we are able to correct a severely under-resolved solution to accurately reflect the long time statistics of the fully resolved reference -- even when the model is trained on much shorter time histories than the reference. Finally, we apply our framework to a realistic climate model, the Energy Exascale Earth System Model (E3SM) with $\sim 110$ km grid resolution. We show that using only 8 years of training data our correction operator is able to bring the global and regional 30-year statistics of the primitive variables into good agreement with ERA5 reanalysis data, and reduce the error in the 36-year average integrated vapor transport (IVT) by 51$\%$ relative to the free-running E3SM solution. Our results show that our framework is able to characterize statistics of events with a return period that is multiple times longer than the length of the training data and therefore represent a promising step towards reliable long term climate predictions.

The remainder of the article is organized as follows. In \S \ref{sec:framework} we introduce the mathematical framework and general machine learning strategy. We then apply our method to a quasi-geostrophic model in \S\ref{sec:QG} and the E3SM climate model in \S\ref{sec:E3SM}. Finally we conclude with a discussion of the implications of our results and the potential extensions and limitations of our method in \S\ref{sec:discussion}.


\section{Training correction operators for imperfect chaotic systems}\label{sec:framework}

We  consider a \textcolor{black}{high-resolution} discretization of an ergodic chaotic dynamical system, and its solution (named thereafter the reference solution), 
\begin{equation}
 \dot {\mathbf{ u}}=F( \mathbf u), \ \  \mathbf u\in\mathbb{R}^N\ \label{perfect_m}
\end{equation}
as well as, a coarse discretization of the same dynamical system (referred as CR), described by the   model\begin{equation}
\dot v= f(v), \ \  v\in\mathbb{R}^n,\label{imperfect_m}
\end{equation}
where $n<N$. The reference solution is projected to the coarse grid through the projection operator $\mathcal P$, i.e.
\begin{equation}
	u=\mathcal{P \mathbf u},\ \   u \in \mathbb R^n
\end{equation}
The objective of this work is to capture the long time statistics of $u$ by solving the imperfect model  (\ref{imperfect_m}) and then applying a correction operator, $\mathcal G$, to the computed solution. The correction operator is assumed to be spatially non-local, with memory, but causal, i.e. the correction at time $t$ may depend only on the past of the input but not the future. To learn this correction operator we assume a reference dataset (referred as RD) generated by the high resolution model or reanalysis data in the form of a finite time trajectory: $\left\{ u(t), \ \ t \in [0,T]\right\}$.

This is a non-trivial problem since any CR trajectory (equation (\ref{imperfect_m})) and RD trajectory (reference dataset $U$) will not be comparable, i.e. cannot be used to formulate the training of the correction operator as a supervised learning problem. In fact, even if the initial condition of the imperfect model is chosen to be the same with $u(t=0)$, the two trajectories will rapidly diverge due to the chaotic \textcolor{black}{nature} of the system. 

In \citeA{blanchard_multi-scale_2022} the authors aim to address this fundamental obstacle by developing a cost function that penalizes directly the deviation between the generated statistics of $\mathcal G (v)$ and the statistics of the reference trajectory, $u$. While the approach has shown some promise, it is a very hard optimization problem that often results in non-physical realizations, $\mathcal G (v)$. At a more fundamental level, the approach does not really utilize the `sequencing' or dynamics encoded in the reference data, but rather its statistics, which for real world problems cannot be guaranteed to be accurate especially for rare events (e.g. using 40 years of reanalysis data cannot guarantee accurate statistics for rare events with a longer return period).

Here we follow a radically different method that aims to learn the correction operator $\mathcal G$ using the reference trajectory and the dynamics of the coarse model, rather than their corresponding finite-time statistics. \textcolor{black}{One of the key objectives of this work is the identification of a dataset which will allow for the training of such a correction operator.  The primary challenge therein is the need to suppress the chaotic  divergence of the coarse scale model during the training phase.} 

We consider the deviation of the two dynamical systems:
  \begin{equation}
q \equiv v-u, \ \ \ q\in\mathbb{R}^n. \label{trans_linear}
\end{equation} 
By computing the derivative we have an equation along the reference trajectory, $\mathbf u$,\begin{align}
\dot q &=f(v)-\mathcal P F(\mathbf u) =f(q+\mathcal P \mathbf u)-\mathcal PF(\mathbf u). \label{div_eq}
\end{align}
The right hand side expresses, for a given  $\mathbf u$, the way the two models diverge. Naturally, the above equation will provide useful information between the two trajectories for as long these remain close to each other. Beyond that point, i.e. after chaotic divergence has occurred, it is not meaningful to compare the two trajectories. To address this issue, we add a damping term in the right hand side of eq. (\ref{div_eq}) that will keep the deviation small:
\begin{align}
\dot q_\tau &=f(q_\tau+\mathcal P \mathbf u)-\mathcal PF(\mathbf u)-\frac{1}{\tau}q_\tau, \label{div_eq1}
\end{align}
where $\tau$ is a constant \textcolor{black}{relaxation time scale} that is chosen so that the added term is at least one order of magnitude smaller compared with all the other terms in (\ref{div_eq1}). Moreover, we add the subscript $\tau$ to emphasize that this is divergence computed with the artificial damping term. The added term is large enough to guarantee that over time scales \textcolor{black}{longer than $\tau$} the deviation does not grow exponentially due to chaotic effects, i.e. the coarse scale model remains in a relevant state to the reference state, but also small enough to allow for the coarse scale model \textcolor{black}{dynamics to evolve unimpeded}. The last point is essential in order to obtain a dataset with sufficient content regarding the imperfection of the coarse scale model. 

By transforming the equation for $q_\tau$ into the $v$ variable, we obtain the final equation for the generation of \textit{nudged} datasets to be used for training:
\begin{align}
\dot v_\tau &=f(v_\tau)-\frac{1}{\tau}(v_\tau-u), \label{div_eq2}
\end{align}
where the second term on the right hand side is known as the nudging tendency. The pair of trajectories $(v_{\tau},u)$ is the basis for training the correction operator. We note that nudging has been widely used in the context of data-assimilation to improve the predictive capabilities of climate models \cite{storch_spectral_2000,miguez-macho_regional_2005,sun_impact_2019,huang_development_2021} as well as on developing hybrid approaches for climate modeling \cite{bretherton_correcting_2022}. Here the use of nudging is only for the development of relevant training pairs of trajectories.

\subsubsection*{Interpretation of training with data from the nudged model}
\textcolor{black}{To obtain a dynamical understanding of the mapping process between the nudged trajectory generated by the above equation and the exact trajectory, we hypothesize the existence of a slow-fast decomposition for $v_\tau$ and $u$. Our motivation is the observation that for many turbulent systems, spatially-coarse modeling affects primarily the fast time scales while it results in smaller errors in the slow time scales. However, fast time scales are important for the characterization of extreme events, as the latter are typically short lived structures.
We express the solution $v_\tau$ in the following slow-fast decomposition based on the relaxation time scale $\tau$:
\begin{align}
	v_\tau(t)=v_s(\mathcal T)+v_f(t),
\end{align}
where $\mathcal T=\epsilon t$ is the slow time scale, and $\epsilon=1/\tau < 1$, where $\tau$ is the relaxation time scale. Moreover, we also decompose the reference solution in a slow-fast form:
\begin{align}
	u(t)=u_s(\mathcal T)+u_f(t),
\end{align}
Based on the above, we have by direct calculation:
\begin{align}
	\dot v_\tau(t)=\epsilon v_s'(\mathcal T)+\dot v_f(t,v_s), \ \ \text{where} \ \ v'=\frac{dv }{d \mathcal T}.
\end{align}
Substituting into (\ref{div_eq2}) we obtain
\begin{align}
\epsilon v_s'+\dot v_f=f(v_s+v_f)+\epsilon(u_s+u_f-v_s-v_f).
\end{align}
Separating the slowly evolving terms of order $\mathcal O(\epsilon)$, i.e. the small terms that depend only on $\mathcal T$, we have:
\begin{align}
v_s'=u_s-v_s \Rightarrow v_s(\mathcal T)= \int e^{-(\mathcal T-s)}u_s(s)ds.  \label{slow_eq}
\end{align}
The fast terms on the other hand will give, to zero order:
\begin{align}
\dot v_f=f(v_s(\mathcal T)+v_f)+\mathcal O(\epsilon).\label{fast_eq}
\end{align}
From the last two equations we can conclude that equation (\ref{div_eq2}) essentially drives the coarse scale model along the slow dynamics of the reference attractor captured by the trajectory, $u$, (\ref{slow_eq}), but leaves the fast dynamics free to evolve according to (\ref{fast_eq}). By driving the imperfect model in regions of the attractor where we have reference data we are able to define a supervised learning problem, where the input is the solution with imperfect fast dynamics defined by (\ref{div_eq2}) and the output is the reference solution, $u$.  In this way, one can use this pair of data to machine learn a map that corrects the fast features of the imperfect model, where the largest model errors are concentrated due to coarse discretization.\\
It is important to emphasize that the method does not assume any scale separation in the dynamics. Instead the parameter $\tau$ controls which temporal scales are corrected by the NN operator. On the other hand, it is important to mention that the success of the scheme relies on a minimum data requirement, sufficient to guarantee proper generalization of the correction operator.}

\subsubsection*{Selection of the relaxation time scale $\tau$}
One of the key questions in the practical implementation of this framework is the choice of the relaxation timescale $\tau$. It quantifies the strength of the nudging tendency and represents a trade off between the suppression of the chaotic divergence and the suppression of the inherent dynamics of the coarse model. If $\tau \rightarrow \infty$, the nudging tendency in (\ref{div_eq2}) will be too weak to suppress the chaotic divergence between $v_{\tau}$ and $u$. This will mean that a map between them will not generalize when applied to free-running coarse solutions. Alternatively, if $\tau \rightarrow 0$, the nudging tendency will completely suppress the dynamics and $v_{\tau}$ will be indistinguishable from $u$ and a map between them will be trivial. From numerical experiments we performed, we found that a value of $\tau$ that results in a nudging term that is one order of magnitude smaller than the other terms of the model represents a good balance between these extremes, i.e. the performance of the algorithm remains the same as long as the choice of $\tau$ remains within this range.

\subsubsection*{Spectrum-matched nudging}
Before we proceed to the machine learning of the correction operator we need to address an energetic inconsistency created by the inclusion of the nudging term in the coarse scale model. This is associated with the artificial dissipation that is introduced to the dynamics of the model due to the term $\frac{1}{\tau}v_{\tau}$. While the term is generally smaller than all other terms of the model, it still creates small discrepancies between the spectra of the nudged solution, $v_\tau$, and the free coarse solution, $v$. This is an inconsistency that has been observed in different settings of data-assimilation and several solutions have been proposed, e.g. 4DVar \cite{mons_reconstruction_2016} or ensemble variational method \cite{mons_reconstruction_2016, buchta_observation-infused_2021}.

\begin{figure}
    \centering
    {\includegraphics[width=0.85\textwidth]{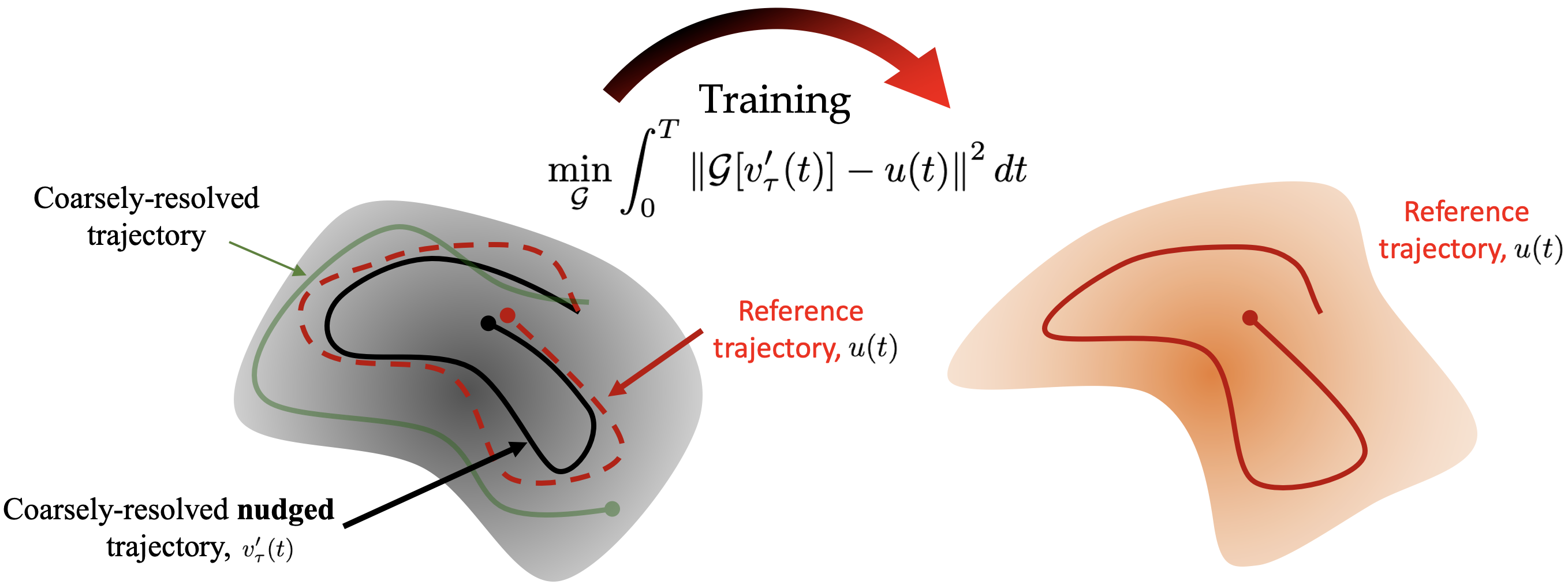} }
    \caption{\textbf{Description of the method } that learns a map between the attractor of the coarsely-resolved equations and the attractor of the reference trajectory. Left: the red dashed curve represents the reference trajectory. The black curve is a coarsely-resolved nudged trajectory towards the reference trajectory. The green curve is the free-run coarsely-resolved trajectory that is not used for training (shown for reference). Right: the target attractor and the target trajectory (red), same as the dashed curve shown at the left plot.}%
    \label{fig:Attractor_Plot}%
\end{figure}

\begin{figure}
    \centering
    \includegraphics[width=0.75\textwidth]{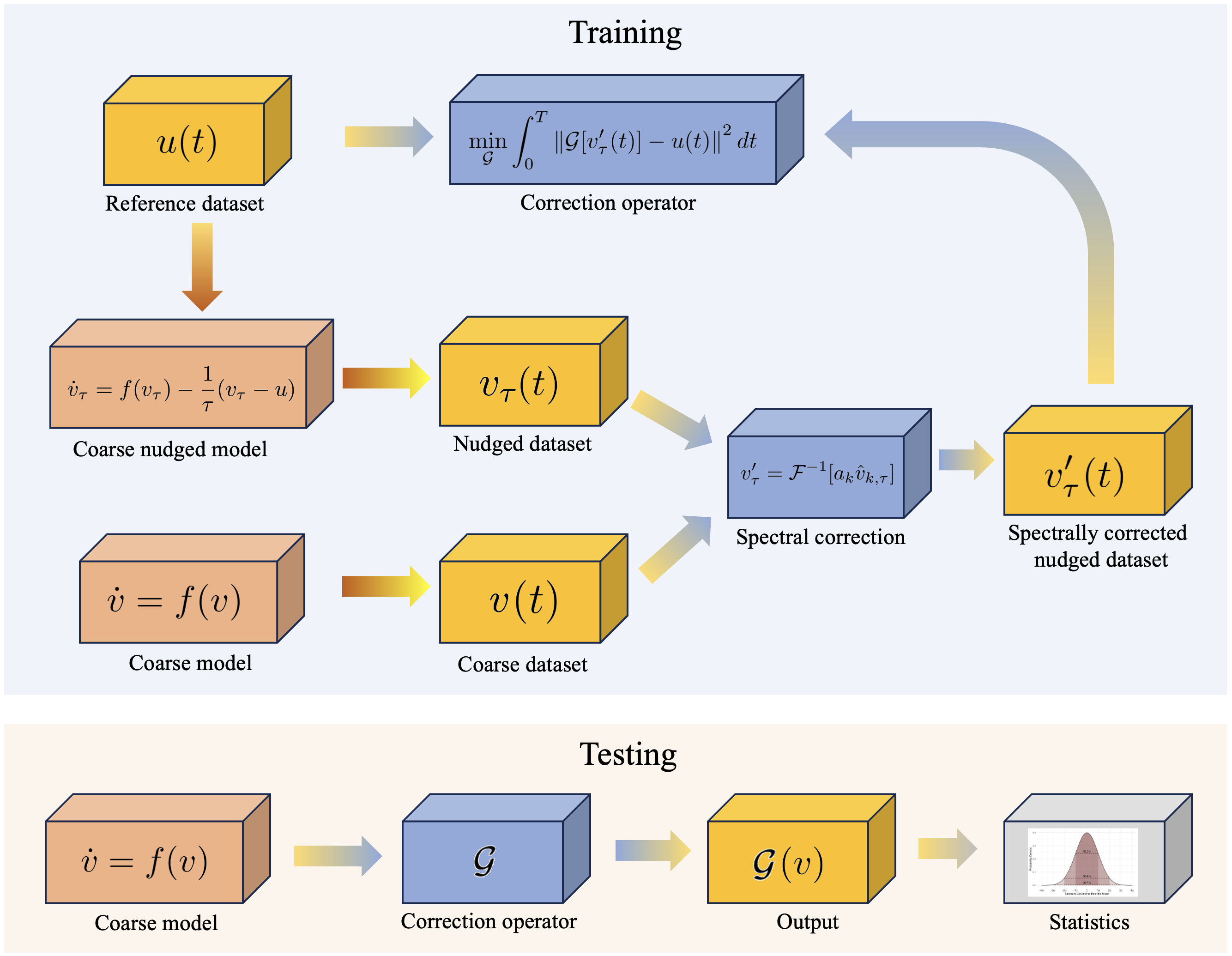} 
    \caption{Workflow of the training process (top) and testing process (bottom) for the machine learning of correction operators and their application on the generation of long time climate simulations, i.e. longer than the reference dataset.}%
    \label{fig:LSTM_Generic_Methodology}%
\end{figure}

Here we employ the simplest approach to correct the spectral inconsistency: we rescale the spectrum of the nudged trajectory, $v_\tau$ to match the spectrum of the coarse model spectrum. Specifically, let $\hat u_k=\mathcal F[u]$ be the spatial Fourier transform of the field $u$. We define the spectral energy as 
\begin{equation}
    \mathcal{E}_{{k,u}} = \frac{1}{T}\int_0^T|\hat{{u}}_{{k}}|^2dt.
\end{equation}
Next, we consider the energy-ratio per wavenumber, between the free-running, $v$, and the nudged solution, $v_\tau$, defined as 
\begin{equation}
    a_{{k}} \equiv \sqrt{\frac{\mathcal{E}_{{k,v}}}{\mathcal{E}_{{k,v_\tau}}}} 
\end{equation}
We define as the spectrum-matched nudged solution as the inverse Fourier transform of the spectrally rescaled nudged solution:
\begin{equation}
    v_\tau'=\mathcal F ^{-1}[a_k \hat v_{k,\tau}].
\end{equation}
The resulted pair of \textit{spectrally-corrected nudged }solution, $v_\tau'$ referred in what follows as NC dataset, together with the reference dataset (RD), $u$ define a supervised learning problem with cost function being:
\begin{equation}
     \min_{\mathcal G} \int_0^T \left\|\mathcal{G}[v_\tau'(t)]-u(t)\right\|^2dt
\end{equation}
The training framework is graphically illustrated in Fig. \ref{fig:Attractor_Plot}. In contrast to previous approaches that aim to match the statistics of the transformed output with statistics of a reference trajectory, the above optimization problem encodes directly the dynamics i.e. the time sequencing of the dataset. This property is crucial for better generalization capabilities, i.e. to train with a short dataset and be able to capture statistics that correspond to much longer simulations. 
After we have machine learned the correction operator, $\mathcal G$, we apply it to the free running coarse model trajectory (CR), $v(t)$. The result is then used to compute statistics and other properties of interest. The workflows for training and testing are summarized  in Fig. \ref{fig:LSTM_Generic_Methodology}. We emphasize that nudging and reference data are used only in the training phase. At the testing phase, the model is using only free-running coarse data and transform it to obtain the correct statistics. The good generalization capabilities of the correction operator allows for its application on much longer time series than those used for training, i.e. the characterization of extreme events with return period that is longer than the training dataset.

\section{Quasi-Geostrophic Model}\label{sec:QG}
\subsection{Background}
As a first example we apply the presented correction method to the two-layer incompressible quasi-geostrophic (QG) flow \cite{qi_predicting_2018}. In a dimensionless form, its evolution equation is given by
\begin{equation}\label{eq:QG_model}
\frac{\partial q_j}{\partial t} + \mathbf{u}_j \cdot\nabla q_j + \left(\beta +k_d^2U_j\right)\frac{\partial \psi_j}{\partial x} = -\delta_{2,j}r\nabla^2\psi_j - \nu \nabla^8q_j
\end{equation}
where $j = 1, 2$ corresponds to the upper and lower layer respectively, $r$ the bottom-drag coefficient and $\beta$ is the beta-plane approximation parameter, and $k_d^2$ represents the deformation frequency which for this study we fix at 4 -- a value consistent with the radius and rotation of the earth and the characteristic length and velocity scales of the atmosphere \cite{qi_predicting_2018}.  This model is intended to approximate mid to high latitude atmospheric flows subject to an imposed shear current. A Taylor expansion of the Coriolis force reveals that for this assumption to hold we require roughly that $\beta \in [1,2]$, which corresponds to an approximate latitude range of $\phi_0 \in [29^{\circ},64^{\circ}]$. 

The flow is defined in the horizontal domain $(x, y) \in [0, 2\pi]$ and is subject to doubly periodic boundary conditions. The state variable is represented in three forms: velocity: $\mathbf{u}_j$, potential vorticity (PV): $q_j$ and the stream function: $\psi_j$. The latter are related via the inversion formula
\begin{equation}
q_j = \nabla^2 \psi_j + \frac{k^2_d}{2}\left(\psi_{3-j} - \psi_j\right)    
\end{equation}
and the velocity is related to the the stream function by
    $\mathbf{u}_j  = U_j +  \hat{\mathbf{k}} \times \nabla \psi_j $
where $\hat{\mathbf{k}}$ is the unit vector orthogonal to the $(x,y)$ plane and $U_j = -1^{(j+1)}U$, with $U=0.2$ represents the imposed mean shear flow. The corresponding nudged system of equations is given by
\begin{equation}\label{eq:QG_model_nudged}
\frac{\partial q_j}{\partial t} + \mathbf{u}_j \cdot\nabla q_j + \left(\beta +k_d^2U_j\right)\frac{\partial \psi_j}{\partial x} = -\delta_{2,j}r\nabla^2\psi_j - \nu \nabla^8q_j - \frac{1}{\tau}\left(q_j - q_j^{RD}\right)
\end{equation}
where $q_j^{RD}$ is the reference solution projected to the grid of $q$. We fix the nudging parameter $\tau = 16$ -- a value for which we found the nudged solution tracks the reference, but generally retains the spectral properties of the free-running coarse solution. Furthermore, we note that while the nudging penalty is applied to the vorticity, it could have equivalently been applied to the stream function or velocity. These possibilities are not explored in this work, however, as these three variables are all directly related we would not expect significant differences in the results.

The equations (\ref{eq:QG_model}) and (\ref{eq:QG_model_nudged}) are solved using a spectral method, with a spectral resolution of $24\times 24$ and $128\times128$ for the coarse- and fine-scale data respectively. The time integration is evaluated using a $4^{th}$ order Runga-Kutta scheme with the same temporal resolution used for both the under- and fully-resolved simulations. Throughout the following discussion all results will be presented in the form of the stream function -- as this uniquely defines the velocity and thus vorticity, this choice incurs no loss of generality. Additionally, we define the zonally averaged stream function as the integral over the $x$ dimension,
\begin{equation}\label{eq:psi_hat}
    \Bar{\psi}_j(y,t) = \frac{1}{2\pi}\int_0^{2\pi} \psi_j(x,y,t) dx.
\end{equation}

\begin{figure}
    \centering
    \includegraphics[trim =  0 0 0 0, width=0.95\textwidth]{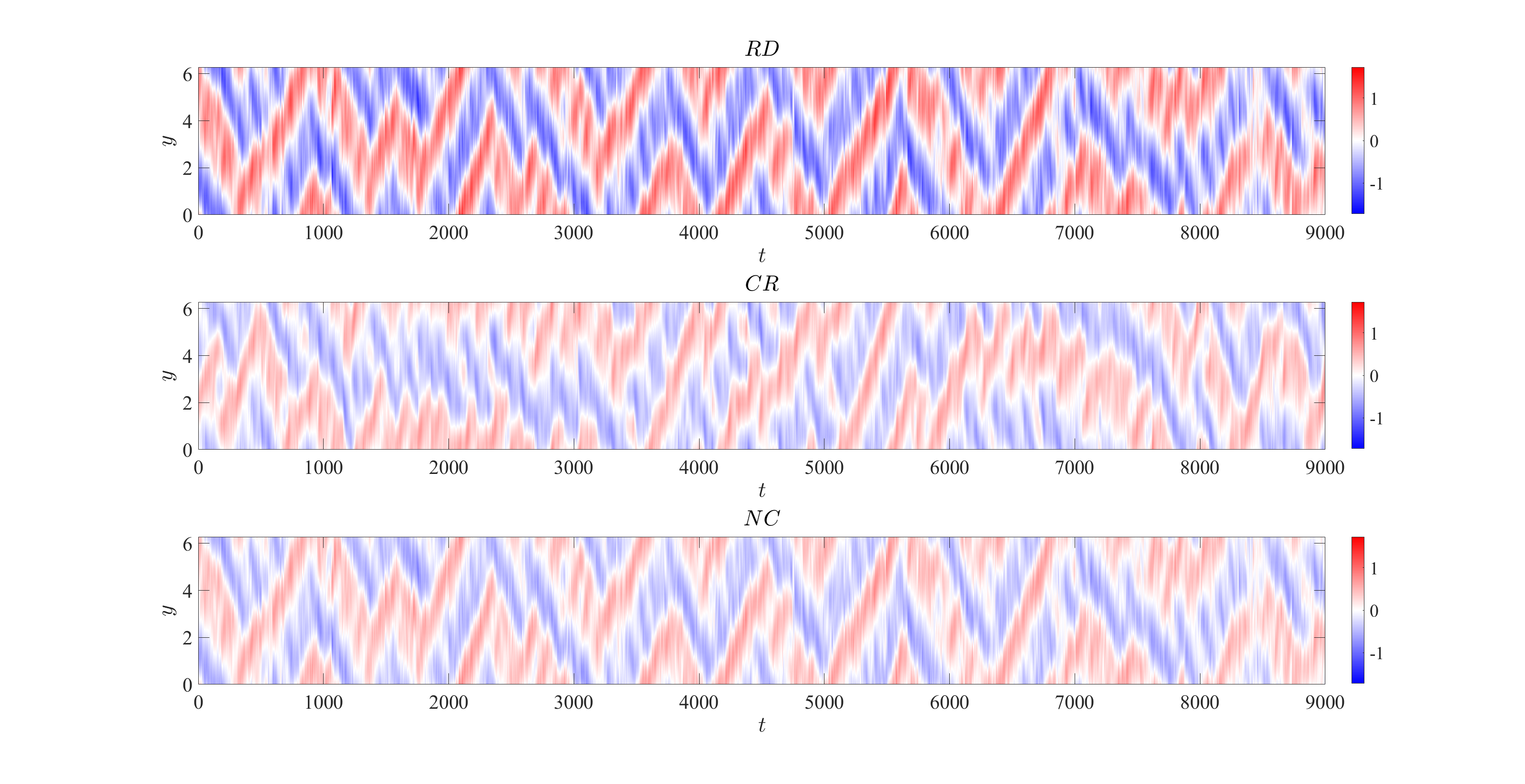} 
    \caption{Example zonally averaged stream function $\hat{\psi}_1$ of the QG system (\ref{eq:QG_model}) for $\beta = 2.0$ and $r = 0.1$. From top to bottom: fully resolved, i.e. reference solution (RD), free-running coarse simulation (CR), spectrally corrected nudged simulation (NC). }
    \label{fig:QG_example}
\end{figure}

In figure \ref{fig:QG_example} we show the zonally averaged stream function in layer 1 for $\beta = 2.0$ and $r = 0.1$ of the three data sets: RD, CR, NC, as an illustrative example of both the fully- and under-resolved solutions. The primary qualitative difference between the coarse and fine grid solutions is in their amplitude. This is particularly clear when comparing the tails of the distributions in \ref{fig:QG_example}b. Note that the spectrally corrected nudged coarse (NC) solution reflects the qualitative spatio-temporal behavior of the fully resolved (RD) solution but exhibits the lower magnitude of the coarse (CR) solution.

\subsection{Neural network architecture and training strategy}
The neural network model we employ as a correction operator takes in as an input the stream function field of both layers which is of dimension $24 \times 24 \times 2$. This vector is then compressed through a fully connected layer of dimension 60 and then passed through a long-short-term-memory (LSTM)  layer of the same size before being expanded through a second fully connected layer to restore the data to its original size. The fully connected layers utilize hyperbolic tangent activation and the LSTM layer uses a hard-sigmoid activation. The model is trained purely on stream function data and thus the output of the model represents the statistically corrected stream function field.

The model is trained on a semi-physics informed loss function which consists of the $L^2$ norm of the error augmented with a second term which penalizes errors in the conservation of mass.
\begin{equation}\label{eq:L2_loss}
    L = \sum_{j=1}^2 \int_{0}^{2\pi}  \int_{0}^{2\pi} |\psi_j^{ml} - \psi_j^{rd}|^2 dx dy + \sum_{j=1}^2 \int_{0}^{2\pi}  \int_{0}^{2\pi} \psi_j^{ml} dx dy
\end{equation}
Here $\boldsymbol{\psi}^{ml}$ and $\boldsymbol{\psi}^{rd}$ denote the machined learned prediction (i.e. the ML transformation of the nudged dataset) and the reference stream functions respectively. The mass conservation term is derived by noting that the two stream functions are linearly related to the height disturbances of the two layers and that by conservation of volume the integral of all height disturbances must vanish.

\textcolor{black}{The correction operator is trained for 2000 epochs on sequences of 100 data points spanning 10 time units taken from} a single realization of the flow with $\beta = 2.0$ and $r = 0.1$ of length 1,000 time units. \textcolor{black}{We then apply the trained correction operator to a separate (unseen) realization of the flow to generate the following results.} .

\subsection{Results}
\subsubsection{Prediction of long time statistics}
First, we apply our models, which are trained on data with $\beta = 2.0$ and $r = 0.1$, to a new realization of the flow with these same parameters. \textit{A key objective of this work is to compute extreme event statistics for events that have a return period that is longer than the length of the training data}. Therefore, the question is how accurately we can capture the tails with a corrected long realization of the coarse model, when the correction operator has been trained \textcolor{black}{on data that does not accurately the tails}, i.e. data of limited length.

\begin{figure}
        \centering
        \centering
        \begin{tabular}{ll}
        \includegraphics[trim =  0 0 0 0, width=0.95\textwidth]{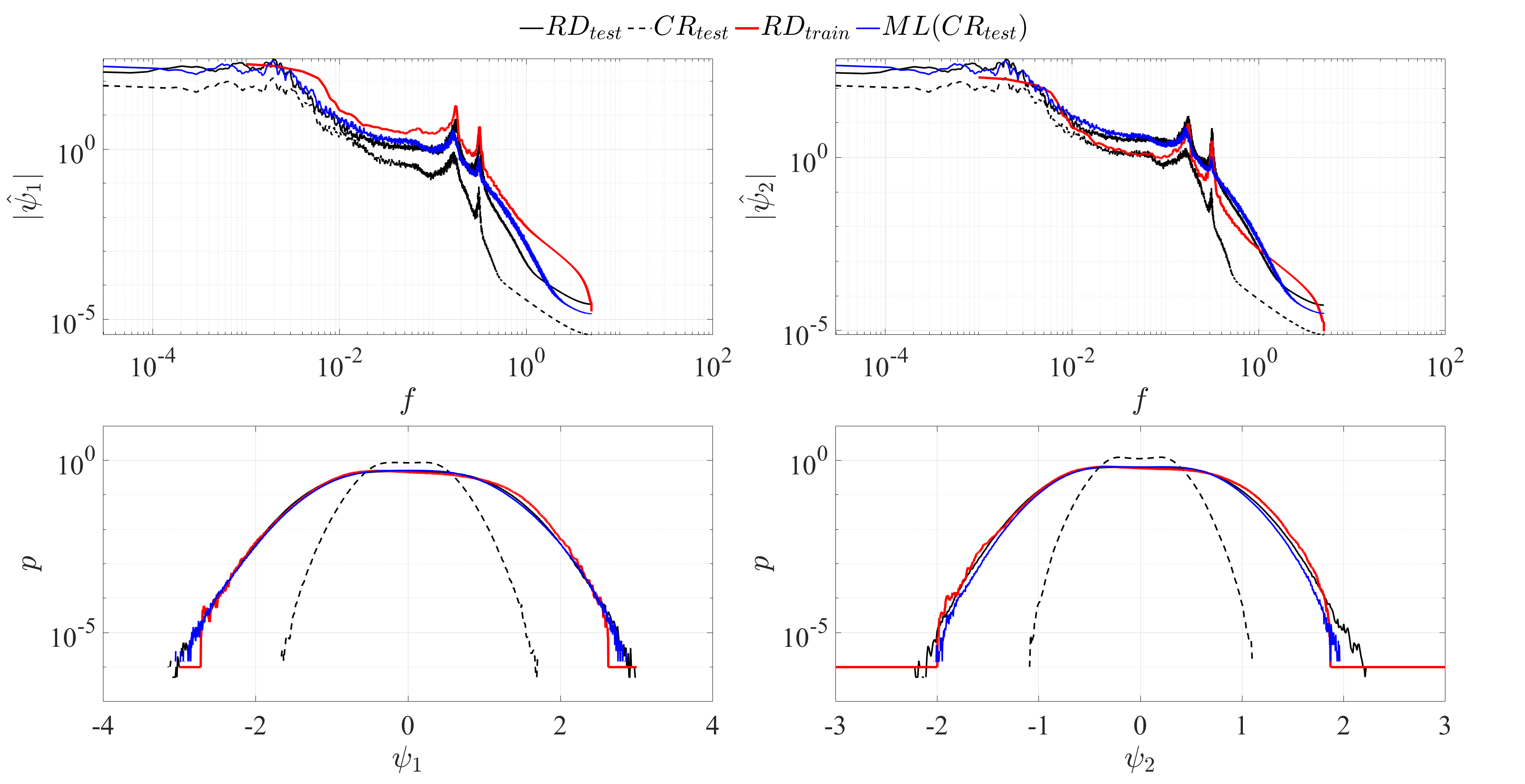} \\
          ~\  ~\   ~\   ~\    ~\  ~\   ~\   ~\   ~\  ~\  ~\  ~\   ~\   ~\   ~\  ~\  ~\  ~\   ~\   ~\   ~\  ~\  ~\  ~\  ~\  ~\  ~\  ~\  ~\  ~\  ~\  (a)  \\
        \includegraphics[trim =  0 0 0 0, width=0.95\textwidth]{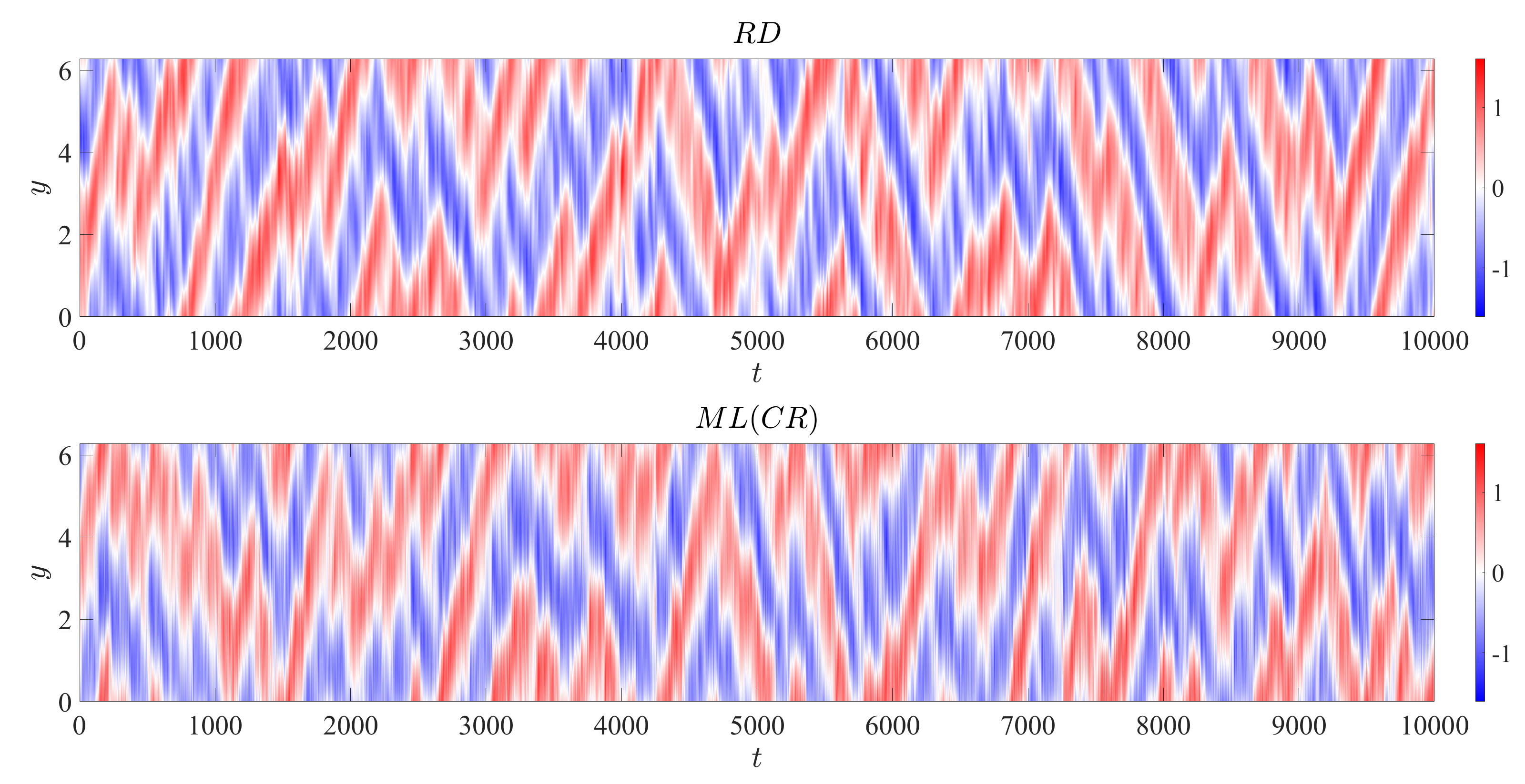} \\
          ~\  ~\   ~\   ~\    ~\  ~\   ~\   ~\   ~\  ~\  ~\  ~\   ~\   ~\   ~\  ~\  ~\  ~\   ~\   ~\   ~\  ~\  ~\  ~\  ~\  ~\  ~\  ~\  ~\  ~\  ~\  (b)   \\
        
        \end{tabular}
        \caption{Model prediction for $\beta = 2.0$ and $r = 0.1$. Power spectrum and probability density function of stream functions $\psi_1$ (top row) and $\psi_2$ (bottom row). Test data, RD (solid black), CR (dash black), ML(CR) (blue) and training data $\mathrm{RD_{train}}$ (red) (a).  Zonally averaged stream function $\Bar{\psi}_1$, RD (upper panel) and ML(CR) (lower panel) (b). $T_{train} = 1,000$ and $T_{test} = 34,000$. }
        \label{fig:QG_results_b2}
\end{figure}

\textcolor{black}{To this end, we} first apply our \textcolor{black}{ML correction operator}, which is trained on \textcolor{black}{$T_{train} = 1,000$ time units of data, to a new realization of the flow spanning $T_{test} = 34,000$ time units}. Figure \ref{fig:QG_results_b2}a shows the global power spectra and probability density functions of the stream function in both layers. The power spectra are computed by taking the spatial average of the point-wise temporal \textcolor{black}{power} spectra, and the probability density function is taken across all space and time.  The fully-resolved (RD) and under-resolved (CR) solutions are shown in solid and dashed black respectively and the ML correction of the under-resolved solution, henceforth denoted ML(CR), is shown in blue. \textcolor{black}{As a reference, we also plot the statistics of the training data ($\mathrm{RD_{train}}$) in red.}

For both layers, the ML correction brings the coarse solution into good agreement with the fully-resolved reference. In terms of the spectra, the ML correction accurately captures the two peaks around $f=0.15$, and only deviates significantly at very high frequencies. In terms of the probability density functions, the model slightly underpredicts the positive tail in layer 2, but captures the general shape well. \textcolor{black}{Crucially, we note that the statistics of the (1,000 time unit) training data are meaningfully different from the (34,000 time unit) test data used to generate the results. Note especially the severe under-resolution of the spectrum and the discrepancy of the far tails of the probability density functions. This highlights the capability of our approach to capture tail events which are not present in the training data, most notably in layer 1. This is an important feature, as any practical long term ($100+$ year) climate prediction will necessarily be trained on far less training data. Furthermore, this highlights the advantages of our approach to one such as \cite{blanchard_multi-scale_2022} in which the ML correction operator is trained to purely reproduce statistics, as such an approach is by construction restricted to the statistics of the training data. }

\begin{figure}
        \centering
        \begin{tabular}{ll}
        \includegraphics[trim =  0 0 0 0, width=0.95\textwidth]{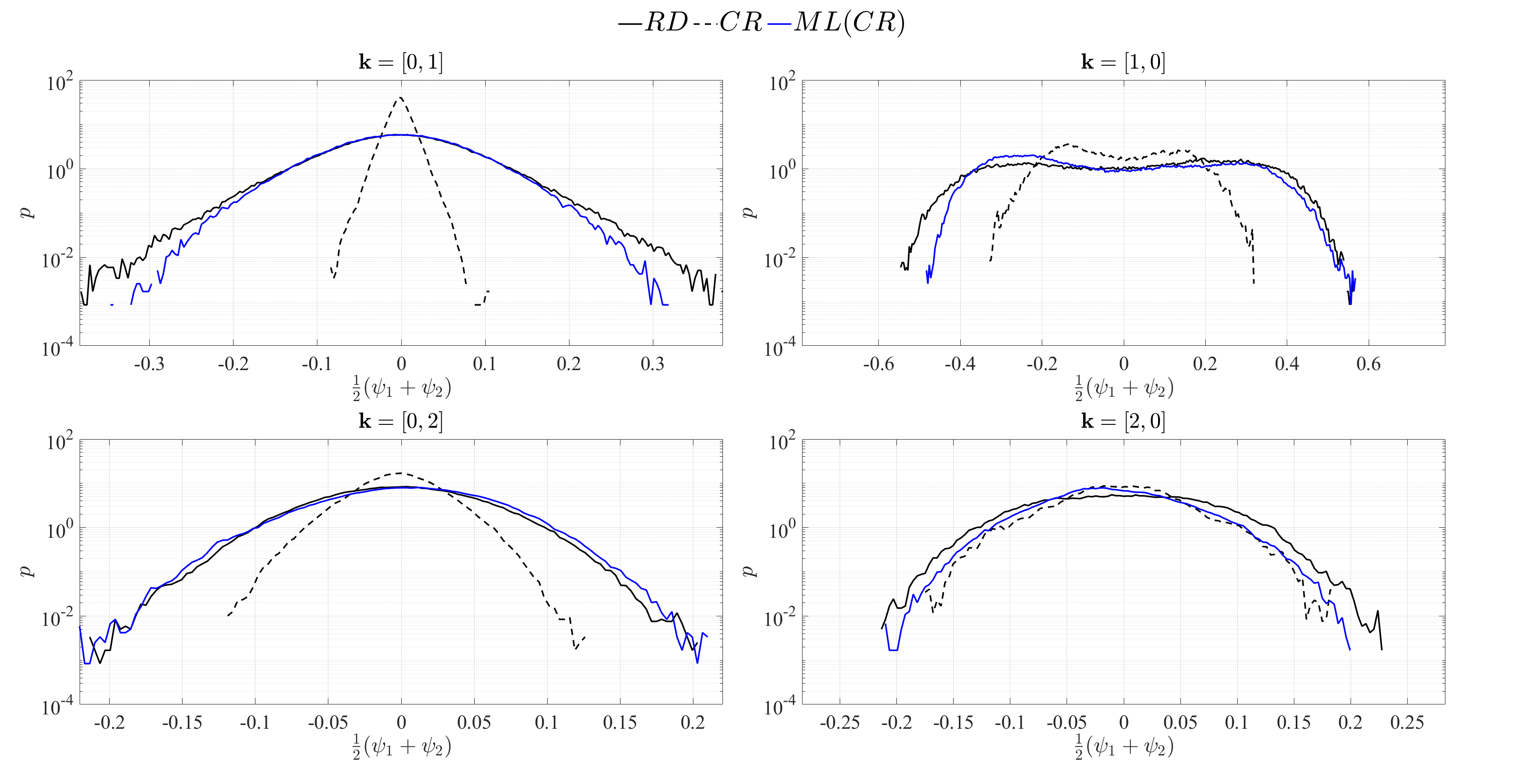} \\
        \includegraphics[trim =  0 0 0 0, width=0.95\textwidth]{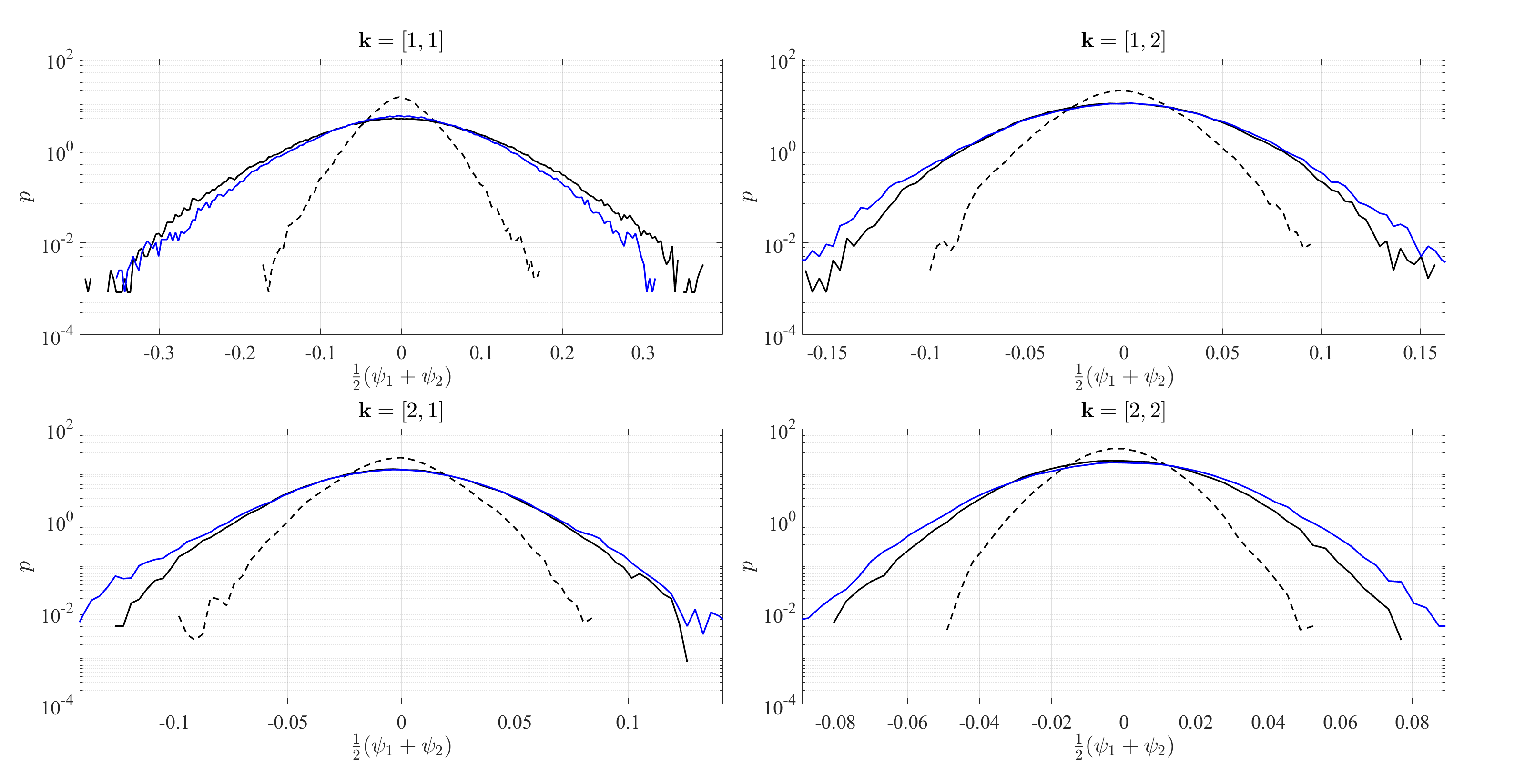} \\
        \end{tabular}
         \caption{Probability density function of individual Fourier modes for $\beta = 2.0$ and $r = 0.1$. RD (solid black), CR (dashed black), ML(CR) (blue) . $T_{train} = 1,000$ and $T_{test} = 34,000$. }
        \label{fig:fourier_pdf2}
\end{figure}

Beyond capturing the global statistics, it is crucial for our model to accurately  capture the dynamics evolving at specific spatial scales. Therefore, in figure \ref{fig:fourier_pdf2} we show the probability density function of a selection of the individual Fourier modes,  parameterized by the wavenumber vector $\mathbf{k} = [k_x,k_y]$.  In the interest of space we show the probability density of the barotropic stream function, defined as the average of the two layers. In general, the model captures the probability distributions of the Fourier modes very well, \textcolor{black}{with some discrepancy in the far tails. Interestingly, the ML correction tends to underestimate the tails of the largest modes e.g. $\mathbf{k}=[0,1]$, and $[1,0]$, while then trending towards overestimating the tails of the smaller modes e.g. $\mathbf{k}=[2,1]$, and $[2,2]$.}

Finally, we reiterate that the only claim we make upon the trajectories predicted by our model is that they reflect the statistical properties of the fully resolved system. However, we expect our predictions to exhibit the qualitative behaviour of the exact solution. To this end we show in figure \ref{fig:QG_results_b2}b \textcolor{black}{a 10,000 time unit interval of }the zonal average of the predicted solution. \textcolor{black}{We do not show the full 34,000 time unit time horizon in order to improve the readability of the figure and highlight the spatiotemporal structure of the flow.} We do indeed find good qualitative agreement with the fully-resolved simulation \textcolor{black}{across the full test trajectory}.


\subsubsection{Minimum training data requirement}
\begin{figure}
    \centering
        \includegraphics[trim =  0 0 0 0, width=1\textwidth]{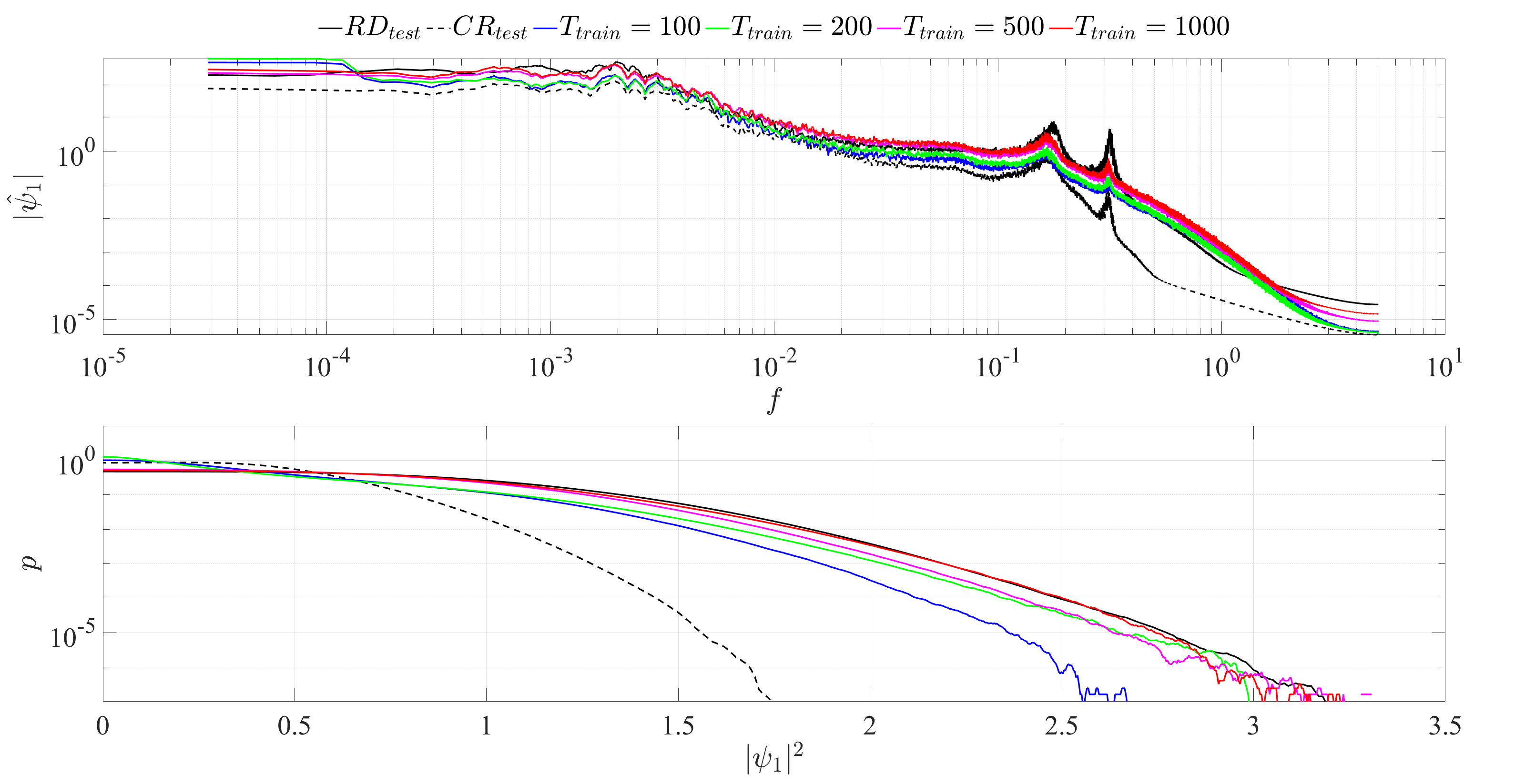}    
    \caption{Model prediction of power spectrum and probability density function of $|\psi_1|$ for $T_{train} = 100$, $200$, $500$, and $1,000$. For all cases $T_{test} = 34,000$.}
    \label{fig:stats_train_times}
\end{figure}

\textcolor{black}{In the previous section we showed that our ML operator is capable of correcting the tails of a long time horizon coarse solution even when trained on a far shorter span of data. Here we investigate the minimum amount of training data needed to capture the long time ($T_{test} = 34,000$ time unit) statistics. We compare the results of our ML correction operator trained on data spanning $T_{train} = 100$, $200$, $500$, and $1,000$ time units -- the latter corresponding to the results described above.} Both training and testing is carried out on data with $\beta = 2.0$ and $r = 0.1$. The probability density function and power spectrum of $|\psi_1|$ for these four cases are shown in figure \ref{fig:stats_train_times}. We focus on the probability density function of the absolute value of the stream function in the interest of brevity. \textcolor{black}{We see that the ML operator requires a minimum $T_{train}$ between 500 and 1000. While, the ML operators trained on $T_{train} < 500$ do improve the statistics of the coarse model, they do not capture the tails of the pdf and also underpredict the two spectral peaks.} This is consistent with a closer examination of figure \ref{fig:QG_example} which shows that the characteristic time scale over which the large scale motions of the flow evolve is approximately 500-1000 time units. Thus, for the QG model considered here, the ML operator requires seeing \textcolor{black}{atleast one full characteristic period of the flow in training. However, once it as seen one or two it is capable of learning the general features of the flow and can accurately reproduce statistics over much longer time horizons}. This is a critical observation since for climate models data is always limited in time and the existence of such critical threshold can indeed pave the way for the computation of statistics for events that have return period much longer than the training data.

\subsubsection{Evaluation for different flow parameters than the training data}
Next, we apply \textcolor{black}{the same ML operator} to a realization of the QG model with flow parameters which differ from the training data, namely $\beta = 1.1$ and $r=0.5$. For these parameter choices the flow lacks the characteristic spectral peaks of the $\beta$ and $r_d$ used to train the model exhibiting much more uniform frequency content.  The lack of a dominant (slower) time scale means the flow evolves on \textcolor{black}{faster characteristic time scale} than the training data. These features make this a challenging test case to evaluate the generalizability of our model. \textcolor{black}{Due to the shorter characteristic time scales, and the associated increased computational cost, for this experiment we consider a test data set of length $T_{test} = 10,000$ time units.}

\begin{figure}
        \centering
        \begin{tabular}{ll}
        \includegraphics[trim =  0 0 0 0, width=0.95\textwidth]{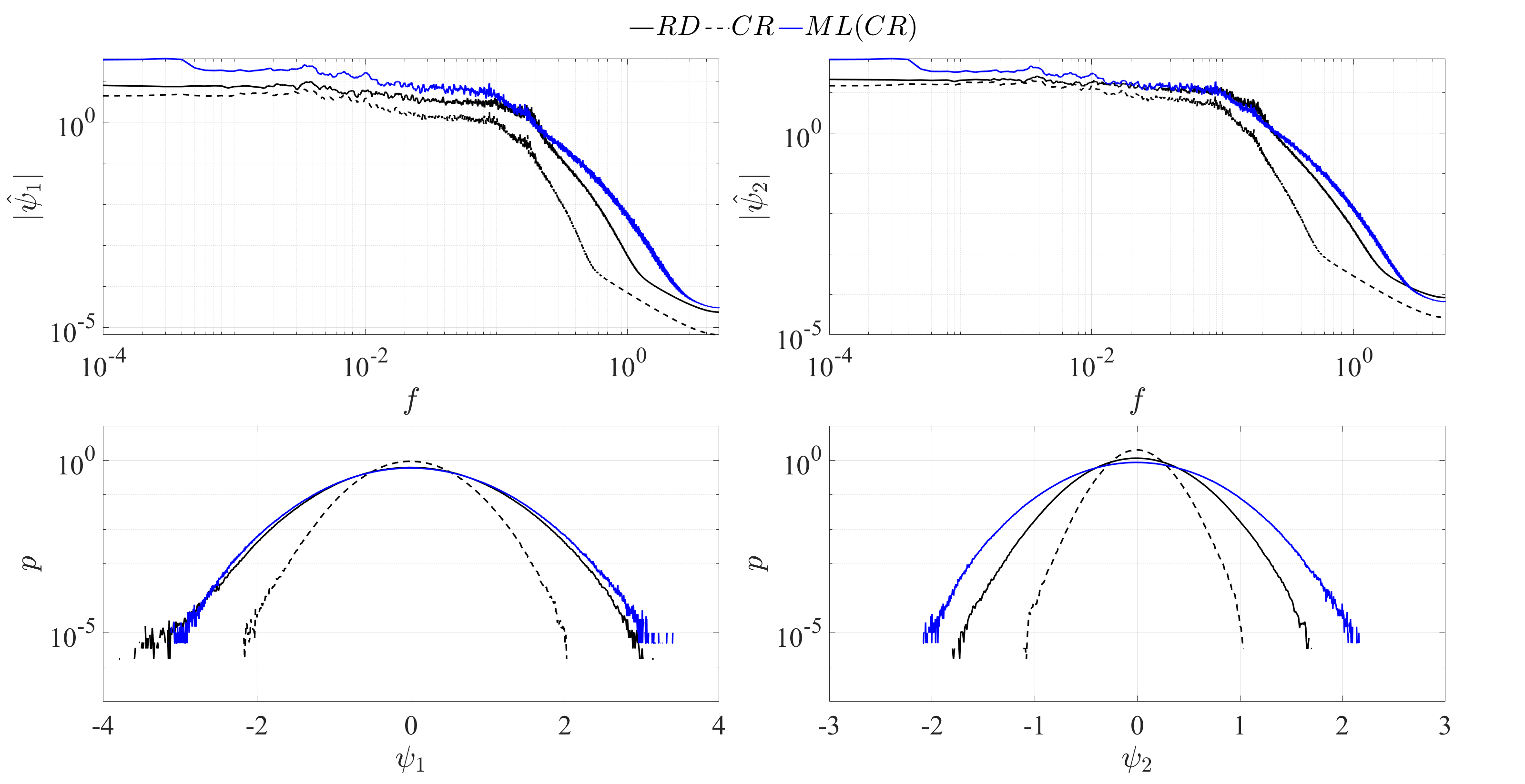} 
        \end{tabular}
        \caption{Model prediction for $\beta = 1.1$ and $r = 0.5$. Power spectrum and probability density function of stream functions $\psi_1$ (left) and $\psi_2$ (right),  RD (solid black), CR (dash black), ML(CR) (blue).  Training data: $\beta = 2.0$ and $r = 0.1$. }

        \label{fig:QG_results_b1}
\end{figure}

\begin{figure}
        \centering
        \begin{tabular}{ll}
        \includegraphics[trim =  0 0 0 0, width=0.95\textwidth]{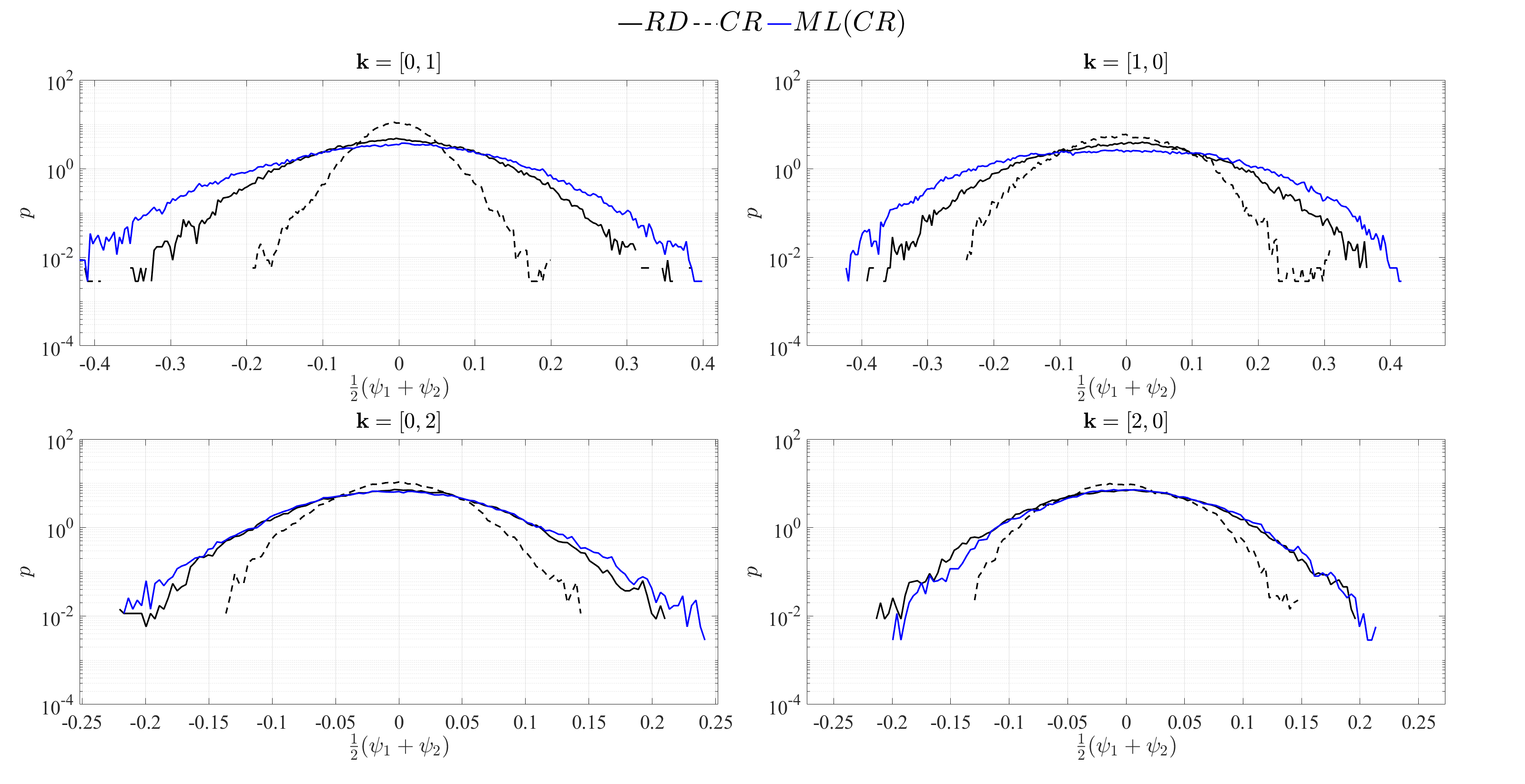} \\
        \includegraphics[trim =  0 0 0 0, width=0.95\textwidth]{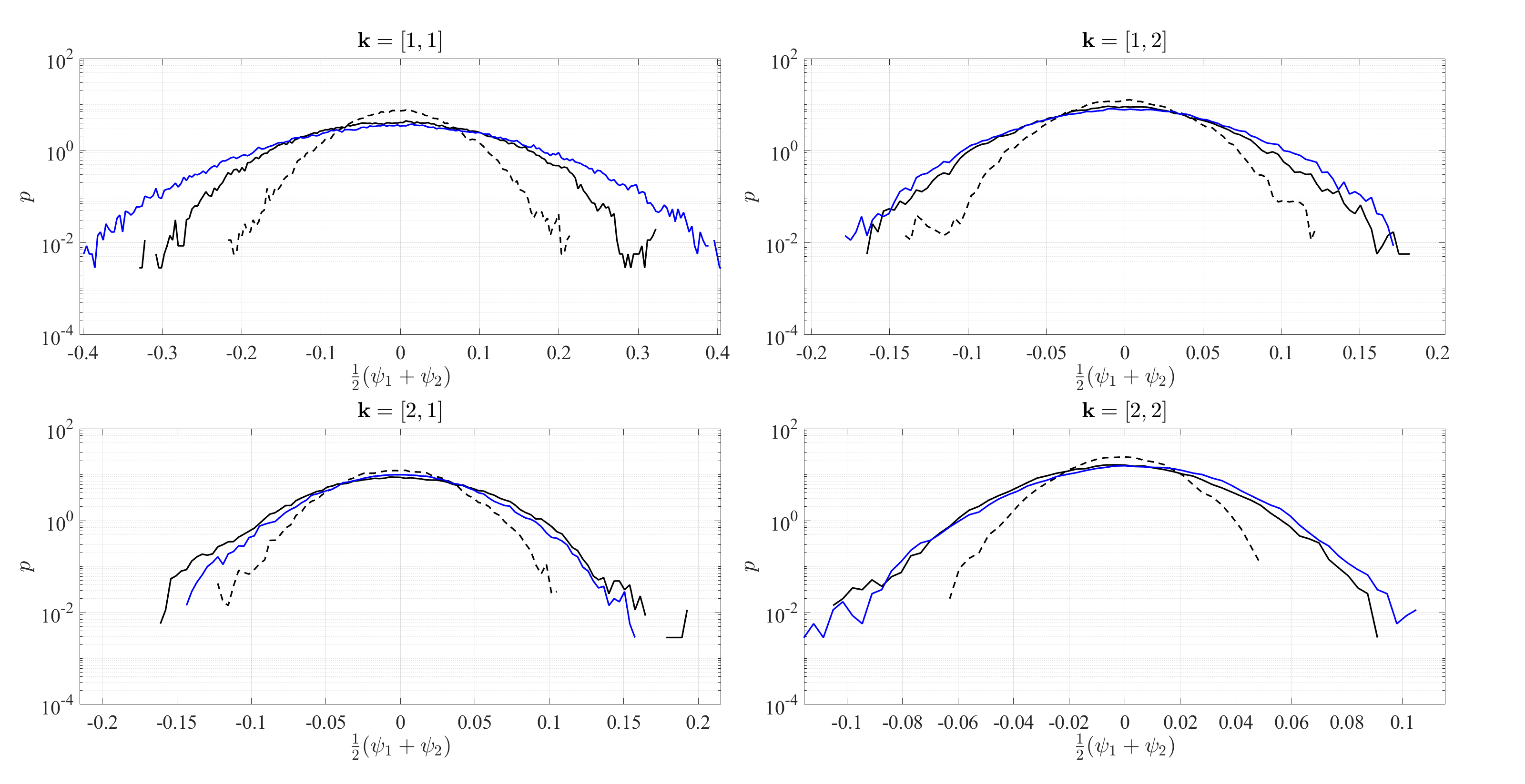} \\
        \end{tabular}
         \caption{Probability density function of individual Fourier modes for $\beta = 1.1$ and $r = 0.5$. RD (black), ML(CR) (blue). Training data: $\beta = 2.0$ and $r = 0.1$.}
        \label{fig:fourier_pdf1}
\end{figure}

The results are summarized in figures \ref{fig:QG_results_b1} and \ref{fig:fourier_pdf1}. In the former we plot the power spectra and probability density function and in the latter we plot the scale-by-scale probability density functions. In terms of the global statistics, the predicted spectrum is in good agreement with the reference across much of the frequency domain, but underpredicts the spectral decay, and thus over-predicts the strength of the highest frequencies. In terms of the probability density function, there is excellent agreement in layer 1, while in layer 2 the model notably over-predicts the tails.  \textcolor{black}{The predictions of the scale-by-scale statistics are reasonably accurate and provide significant improvement over the free-running coarse model. However, the ML correction tends to over emphasize the strength of the tails for the larger length scales, e.g. $\mathbf{k}=[0,1], [1,0], [1,1]$. This is not surprising finding given the drastic over-correction of the tails in layer 2 seen in figure \ref{fig:QG_results_b1}. }


\section{Global Climate Model}\label{sec:E3SM}
\subsection{Dataset}
We now apply our framework to a realistic global climate model, the Energy Exascale Earth System Model (E3SM). In particular, version 2 of the E3SM Atmosphere Model (EAMv2) \cite{dennis2012cam,taylor2009non, golaz2022doe}. The progress variable is $\mathbf{X}\left(\theta,\phi,k,t\right) = \left(U, V, T, Q\right)$. The progress variables $(U,V)$ correspond to the zonal and meridional components of wind velocity, $T$ is air temperature and $Q$ is specific humidity. The spatial coordinates $(\theta,\phi,k)$ are the polar, $\theta \in [-90,90]$,  azimuthal angles, $\phi \in [0,360]$, and the sigma level respectively. \textcolor{black}{The latter of which can be understood as a measure of altitude. We use a hybrid sigma-pressure coordinate system} -- near the surface, the \textcolor{black}{levels} are terrain following, while at higher altitudes they are defined as levels of  constant pressure \cite{taylor2020energy}.
The EAMv2 model pairs the resolved atmospheric dynamical equations with a variety of the sub-grid parameterizations such as cumulus convection \cite{zhang1995sensitivity}, boundary layer cloud dynamics \cite{golaz2002pdf}, cloud micro-physics \cite{morrison2008new}, aerosol micro-physics and chemistry \cite{liu2016description}, and radiative transfer \cite{mlawer1997radiative}.  The coarse-scaled simulations are run on an unstructured \textcolor{black}{spherical element} grid of approximately $1^{\text{o}}$($\sim 110[\text{km}])$ resolution per sigma-level and 72 levels along the vertical direction, from  $64 [\text{km}]$, corresponding to $\sim 0.1 [\text{hPa}]$ (level 1) down to the earth's surface (level 72). The vertical grid spacing is uneven, with the layer height ranging from 20–100 m near the surface up to 600 m in the upper atmosphere. We enforce appropriate boundary conditions over the Earth's surface in accordance with version 4.5 of the community land model \cite{oleson2013technical}. The (SST) and sea ice concentration (SIC) boundary conditions are set according to the input4mip datasets \cite{Reynolds_2002_SST}.

In this case, the reference data used to generate the nudged training data and the validation reference is not a fully-resolved simulation but instead ERA5 reanalysis data \cite{Hersbach_et_al:2020} projected onto the coarse unstructured grid of EAMv2. The ERA5 dataset combines observations with physics models to provide high-quality reanalysis data on an hourly basis with a spatial resolution of $0.25^{\text{o}}$($\sim 31[\text{km}])$. An outline of the practical implementation of the nudging is summarized in appendix \ref{app:nudg_workflow}.

We do not perform any E3SM simulations at this fine resolution due to the prohibitive computational cost, and so in the following discussion any reference to E3SM data should be understood to represent the coarse model. Moving forward, the free-running dataset will again be labeled as CR, the ML correction thereof as ML(CR), and the ERA5 reference data as RD.  The datasets discussed herein contain information from 1979-2014, over which the climate system can be assumed to be in an approximately statistical steady state.




\subsection{Neural network architecture and training strategy}\label{sec:E3SM_NN_Architecture}
For the E3SM model we have developed a custom convolutional-LSTM hybrid network architecture. The architecture acts on a single sigma level, such that training is conducted for each level sequentially. The network receives as its input snapshots of the predictive variables \textcolor{black}{$\mathbf{X} = \mathbf{X}(\theta,\phi,t,k)$} for fixed sigma level $k$. Afterwards, a custom \textcolor{black}{``split''} layer separates the input into \textcolor{black}{25} non-overlapping subregions. These subregions are periodically padded via a custom padding process, tasked with respecting the spherical periodicity of the domain. Then, each subregion is independently passed through a series of \textcolor{black}{four} convolutional layers. The purpose of this process is to extract anisotropic local features in each subregion such as vapor transport. 

\begin{figure}
    \centering
    {\includegraphics[width=0.8\textwidth]{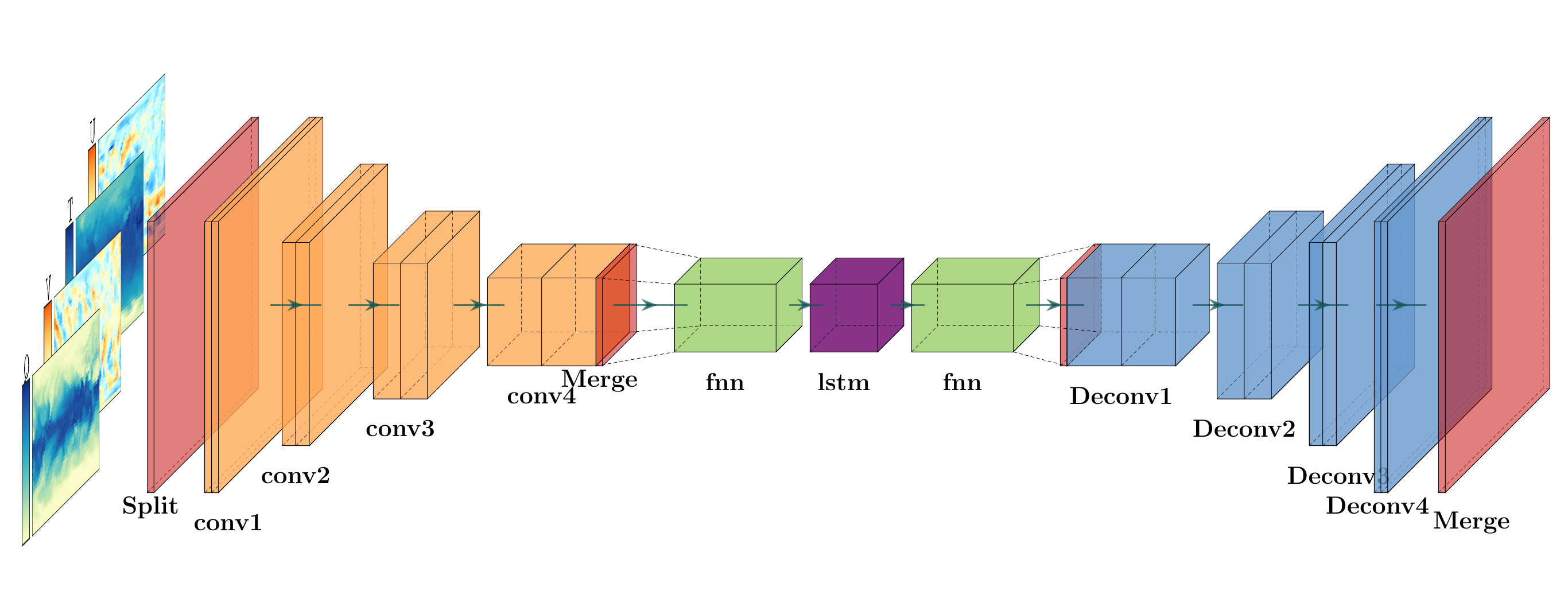} }%
    \caption{LSTM based neural network architecture used for the E3SM climate model.}%
    \label{fig:LSTM_Architecture}%
\end{figure}

Afterwards, the local information extracted from each subregion is concatenated in a single vector via a custom `merge' layer. The global information is now passed through a linear fully-connected layer, that acts as a basis projection of the spatial data onto a reduced-order \textcolor{black}{20-dimensional} latent space. The latent space data are then corrected by a LSTM layer \cite{hochreiter1997long}. Subsequently they are projected back to physical space via another linear fully-connected layer. Next, global information is split into the same subregions of the input, and distributed to \textcolor{black}{another} series of \textcolor{black}{four} independent deconvolution layers that upscale the data to the original resolution. Finally, a custom `merge' layer gathers the information from each subregion and produces the final corrected snapshot. A schematic of the configuration for training on a particular layer is shown in figure \ref{fig:LSTM_Architecture}.
The motivation behind using LSTM neural networks lies in their ability to incorporate (non-Markovian) memory effects into the reduced-order model. This ability stems from Takens embedding theorem \cite{takens1981detecting}. This theorem states that given delayed embeddings of a limited number of state variables, one can still obtain the attractor of the full system for the observed variables. In addition to temporal nonlocality, the model is nonlocal in space. Note, that in terms of the LSTM layer, this information comes in the form of the latent space coefficients, which in general correspond to global modes that correspond to rows of the fully connected layer's matrix. Under the assumption that both fully-connected layers have linear activation functions, the model can be mathematically depicted as a basis projection. Hence, the fully connected layers act as projection schemes to (a) compress input data to a latent space of low dimensionality, and (b) project the LSTM prediction to physical space. Such LSTM based models have been shown to be capable of improving predictions of reduced-order models in a variety of settings  \cite{ vlachas2018data, wan_data-assisted_2018, harlim_machine_2021, charalampopoulos_machine-learning_2022}. However, we note that other network architectures are possible, such as the recently proposed Fourier-Neural operators \cite{li_fourier_2021,li_fourier_2022,guibas_adaptive_2022,bonev_spherical_2023}
which have shown remarkable skill in data-driven weather prediction \cite{pathak_fourcastnet_2022}.

The network is trained using a standard mean-square error (MSE) loss function
\begin{gather}\label{eq:MSE_loss}
    \mathcal{L} = \alpha \sum_t \sum_{\phi} \sum_{\theta} \cos \left( 2\pi \frac{\theta}{360} \right) \lVert \mathbf{X}^{\text{ml}} -\mathbf{X}^{\text{rd}} \rVert^2,
\end{gather}
where $\alpha$ is a normalization coefficient. As previously, training is performed using the nudged dataset as input to the ML transformation. Each term in the sum is multiplied by a cosine that is a function of the latitude to showcase that the integration takes place over a sphere. If that term is absent, the model would over-emphasize on learning the corrections at the poles. Training was conducted over 1000 epochs using data from the years 2007-2011, with the year 2012 used for validation during training.

\subsection{Results}
We apply our model to an unseen free-running coarse-scale simulations of the E3SM model (CR) over a 36 year horizon. These results are denoted as $ML(CR)$. The reference statistics used to evaluate our model predictions are computed from ERA5 reanalysis data over the years 1979-2014 and are denoted as $RD$. We also show the predictions of a free running E3SM simulation denoted $CR$, this serves as the baseline which our model is seeking to improve.

\subsubsection{Global statistics}\label{sec:global_stats}
\begin{figure}
    \centering
    \includegraphics[width=0.95\textwidth]{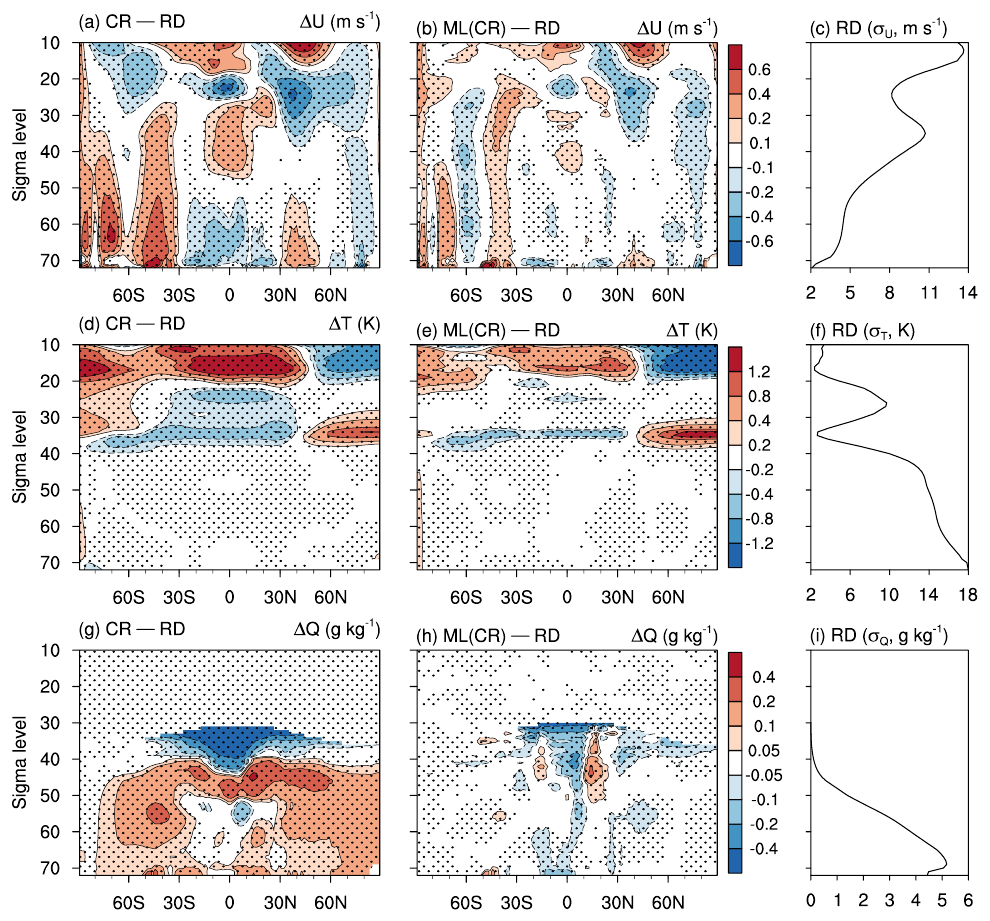}
    \caption{Zonally-averaged 36 year annual mean biases for all sigma-level of the simulations, for normalized zonal velocity $U$ (a-c), temperature $T$ (e-g), and specific humidity $Q$ (i-k).  \textcolor{black}{Free running coarse E3SM simulation (CR) (left) and ML-correction (ML(CR)) (right).} Standard deviation $\sigma$ of each quantity at the specific sigma-level shown  (d,h,i).}
    \label{fig:CLIM_Global_Bias}%
\end{figure}

First, we analyze the global 36-year statistics as a function of altitude, i.e. for all sigma levels. In figure \ref{fig:CLIM_Global_Bias}, we show the time- and zonally-averaged biases for sigma-levels 10-72 of the simulations for $U$ (a-c), $T$ (e-g),  $Q$ (i-k). We omit the highest sigma levels 1-10, as here the reference data is less reliable and thus obscures the analysis. The left column shows the biases of the free-running E3SM while the right column shows those of the ML corrected. The biases are normalized with the standard deviation of the quantity of interest for each sigma-level individually (sub-figures c,f,i). For the case of $Q$ for sigma-levels below $z = 35$, the standard deviation of level $35$ was used for normalization. This is due to the fact that the values of $Q$ in the upper atmosphere are extremely low and normalizing such errors by the standard deviation of their own sigma-level yielded very high biases for both predictions, making the metric misleading. The dotted regions indicate where the biases are statistically significant up to a 95\% confidence level as quantified by a Student-$t$ test. The ML correction notably corrects the strong overestimation of the specific humidity (bottom row) for sigma levels $z>40$. The biases in temperature (middle row) in the upper atmosphere are also notably improved, however the improvement is less pronounced. In the case of the wind speed (top row), the ML correction does reduce the bias throughout the atmosphere, however, both the free running E3SM and the ML correction thereof retain significant biases in the upper atmosphere.

\begin{figure}
    \centering
    \subfloat[]{%
    \includegraphics[width=0.35\textwidth]{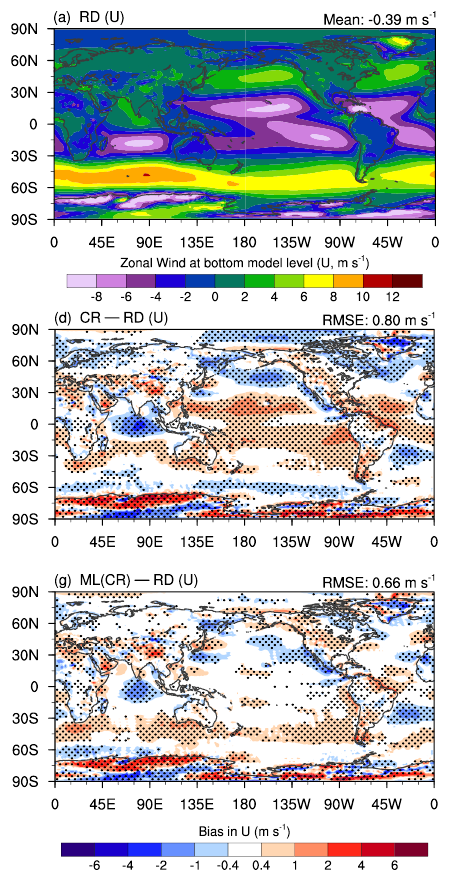}%
}
    \subfloat[]{%
    \includegraphics[width=0.35\textwidth]{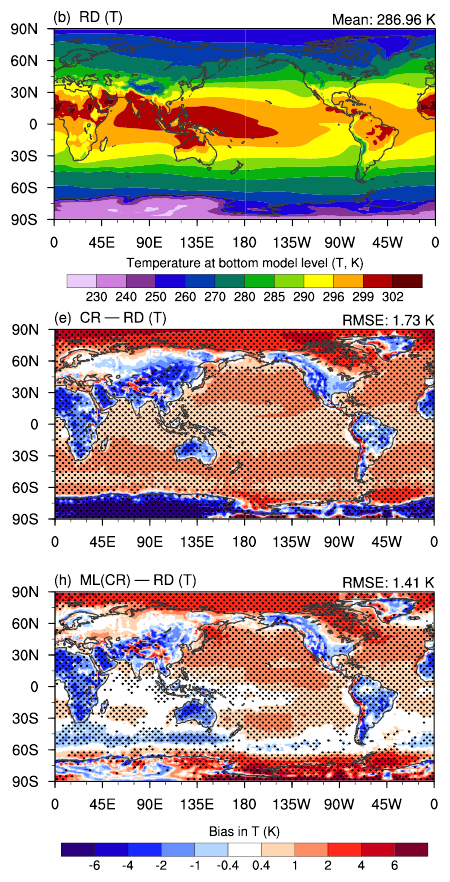}%
}
    \subfloat[]{%
    \includegraphics[width=0.35\textwidth]{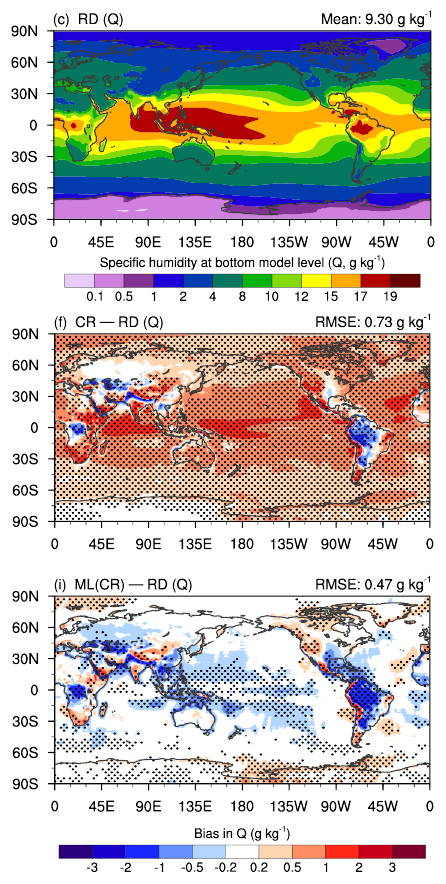}%
}
    \caption{Biases at the lower-most sigma-level with respect to ERA5 for time-averaged zonal velocity $U$, temperature $T$ and specific humidity $Q$. Top row corresponds to the \textcolor{black}{reference data (RD)}, second row corresponds to a free-running simulation (CR) and bottom row corresponds to ML-correction (ML(CR)). }%
    \label{fig:CLIM_Level72_Biases}%
\end{figure}

\begin{figure}
    \centering
    \subfloat[]{%
    \includegraphics[width=0.65\textwidth]{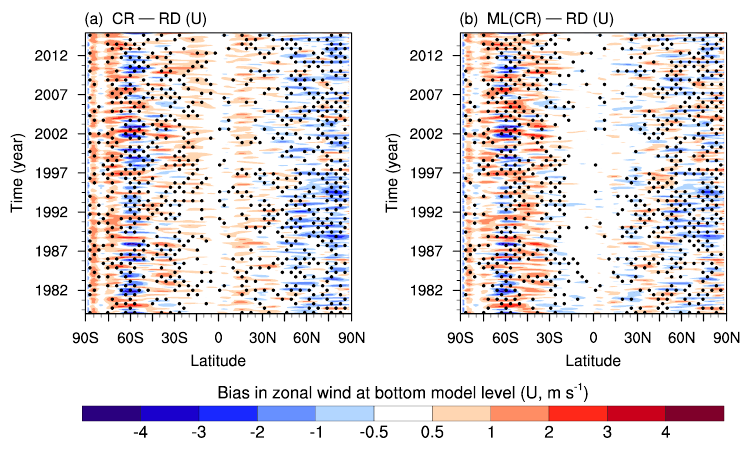}%
}

    \subfloat[]{%
    \includegraphics[width=0.65\textwidth]{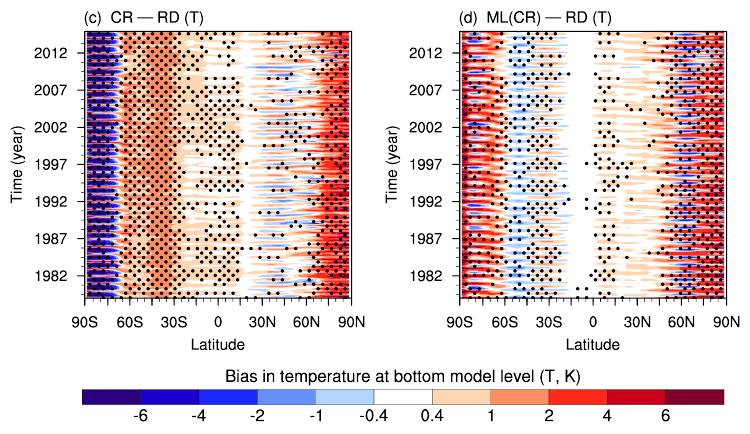}%
}

    \subfloat[]{%
    \includegraphics[width=0.65\textwidth]{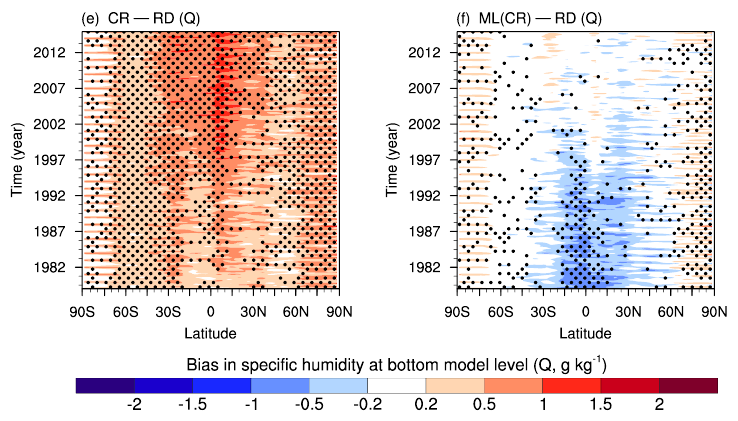}%
}
    \caption{Hovmoller Diagrams of biases at the lower-most sigma-level with respect to ERA5 for time-averaged zonal velocity $U$ (a), temperature $T$(b), and specific humidity $Q$(c). Free running coarse E3SM simulation (CR) (left) and ML-correction (ML(CR)) (right). }%
    \label{fig:Hovmoller_Level72_Biases}%
\end{figure}

We now focus on the sigma level nearest the surface -- level 72. Additional results, including probability density functions over all sigma levels are included in \ref{app:additonal_E3SM_stats}. Figure \ref{fig:CLIM_Level72_Biases} shows the annual mean ERA5 reference data, as well as the biases of the free-running and ML corrected predictions. The ML correction reduces the global RMSE by $18$, $19$, and $36\%$ for $U$, $T$, and $Q$ respectively. Regionally, the benefits of our model correction are best seen in the equatorial and south polar regions. In the former, the free-running solution significantly overestimates the specific humidity, while the ML correction is relatively free of any such systematic bias. Then in the latter, the uncorrected simulation significantly underestimates the temperature, a deficit which is remedied with the ML correction. 
\textcolor{black}{To illustrate the temporal evolution of the near surface biases we also show in figure \ref{fig:Hovmoller_Level72_Biases} the time versus latitude Hovmoller diagrams of the monthly mean zonal mean bias in $U$, $T$, and $Q$ over the time period 1979-2014. We note that the period 2007-2014 is part of our training data. Consistent with the results in figure \ref{fig:CLIM_Level72_Biases}, our ML correction consistently reduces the zonal mean biases of all three quantities. The most significant improvements are observed in T and Q, for which the performance of the ML correction is greatest in the tropical and subtropical regions. Furthermore, in those regions where we observe significant bias reduction, the corrections persist robustly across the years outside the training period. However, there is an over-correction of the positive biases in Q in the tropical regions during the period 1979-2002 (\ref{fig:Hovmoller_Level72_Biases}c). This is possibly because the training data is too short to capture the multi-decade trend of the E3SM model increasingly overestimating the humidity in the tropics}

\begin{figure}
    \centering
    \includegraphics[ width=1\textwidth]{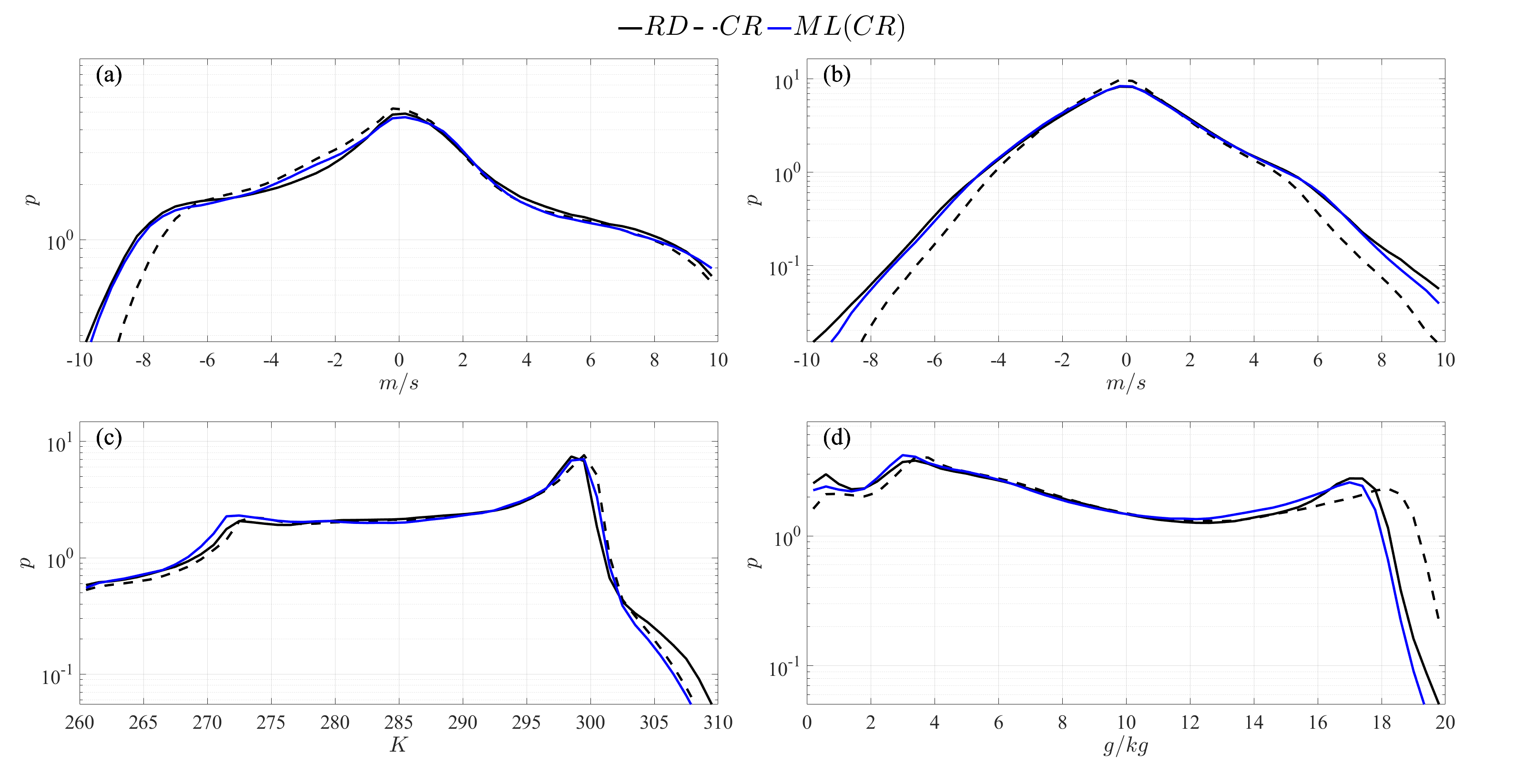}
    \caption{Global 36 year probability density function for surface sigma-level 72.  $U$ (a), $V$ (b), $T$ (c), $Q$ (d). Results are shown for ERA5 reanalysis data (RD) (solid black), free-running data (CR) (dashed black), and ML corrections (ML(CR)) (blue).}
    \label{fig:global_72}
\end{figure}

Figure \ref{fig:global_72} shows the aggregate probability density function at sigma level 72 across the globe for the same 36 year period. The probability density functions are computed using the $36\times12$ monthly mean values at each grid point. The ML correction significantly improves the predicted distributions in wind speed $U,V$ (a, b) and specific humidity $Q$ (d). Critically, the improvements are most pronounced in the tails of the distribution, which are critical for quantifying the risks of extreme weather events. There is very little improvement in the temperature $(T)$, however, in this case the E3SM prediction alone is already quite accurate.

\subsubsection{Integrated Vapor Transport}
We now move to predict statistics for a derived integral quantity, the mean integrated vapor transport (IVT). The IVT quantifies the vertically integrated mass transport of water vapor and is defined as
\begin{equation}\label{eq:IVT}
    IVT(t,\theta,\phi) \equiv \sqrt{IVT_{U}^2 + IVT_{V}^2}
\end{equation}
where $IVT_{U}$ and $IVT_{V}$ are the east-west and north-south components defined as
\begin{equation}
    IVT_{U}(t,\theta,\phi) \equiv \frac{1}{g}\int Q(t,\theta,\phi,p)U(t,\theta,\phi,p) dp
\end{equation}
and similarly for $T_{VQ}$, and where the vertical coordinate has been re-parameterized in terms of pressure. Regions of concentrated IVT are known as atmospheric rivers (AR) and are associated with heavy precipitation and a variety of extreme weather events -- both beneficial and detrimental. For example, on the open ocean, ARs are generally associated with extratropical cyclones, and upon landfall ARs have the potential to alleviate drought conditions or lead to significant storm damage \cite{payne_responses_2020}. Therefore, the ability to correctly predict the statistics of the IVT -- and thus ARs -- is a crucial metric by which to evaluate our ML correction operator. \textcolor{black}{Although it is beyond the scope of this work, the interested reader is referred to \cite{zhang_machine_2023} for a detailed discussion of our method applied to the statistics of other extreme climate events such as tropical cyclones.}

From a machine learning point of view, accurately predicting the spatial features of \textcolor{black}{extreme events, which are quantified by highly anisotropic quantities such as} IVT, requires accurately mapping local flow features between the under- and fully- resolved trajectories. It is for this reason, that we have implemented the domain-splitting and local convolution layers in the network architecture described in  \S \ref{sec:E3SM_NN_Architecture}.

\begin{figure}
    \centering
    \includegraphics[ width=0.65\textwidth]{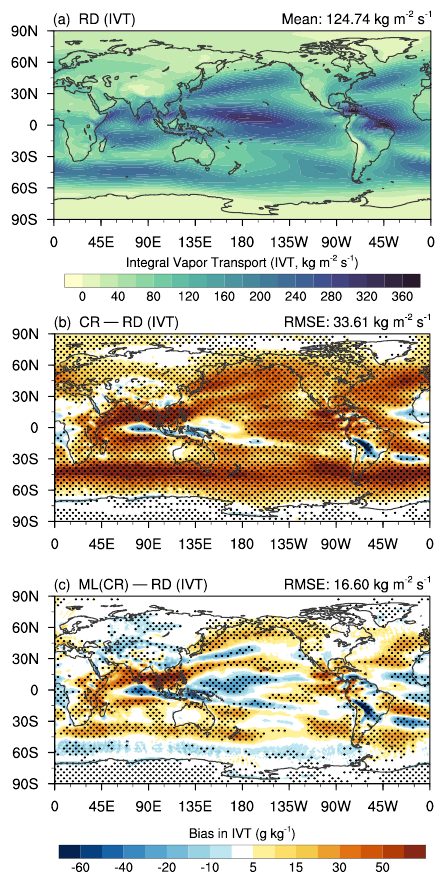}
    \caption{36 year annual mean IVT predictions. From top to bottom, ERA5, free-running E3SM bias, ML correction bias. }%
    \label{fig:AR_IVT_Mean25}%
\end{figure}

In figure \ref{fig:AR_IVT_Mean25}, we show the 36-year annual mean of the integrated vapor transport across the globe. The top figure corresponds to the ERA5 reanalysis data, and below that are the biases of the free-running E3SM simulation, as well as the machine learned correction. Overall, the ML correction decreases the global root-mean-square error (RMSE) by $51\%$ compared to the free-running E3SM solution. Furthermore, the ML correction significantly decreases several systematic regional biases throughout the domain. Note for example, that the ML significantly reduces the strong positive bias of the free-running E3SM simulation over Southeast Asia and in the southern oceans around $45\deg$ of latitude.

\newpage
\newpage

\subsubsection{Regional Statistics}\label{sec:regional_stats}
\begin{figure}
    \centering
    \includegraphics[ width=1\textwidth]{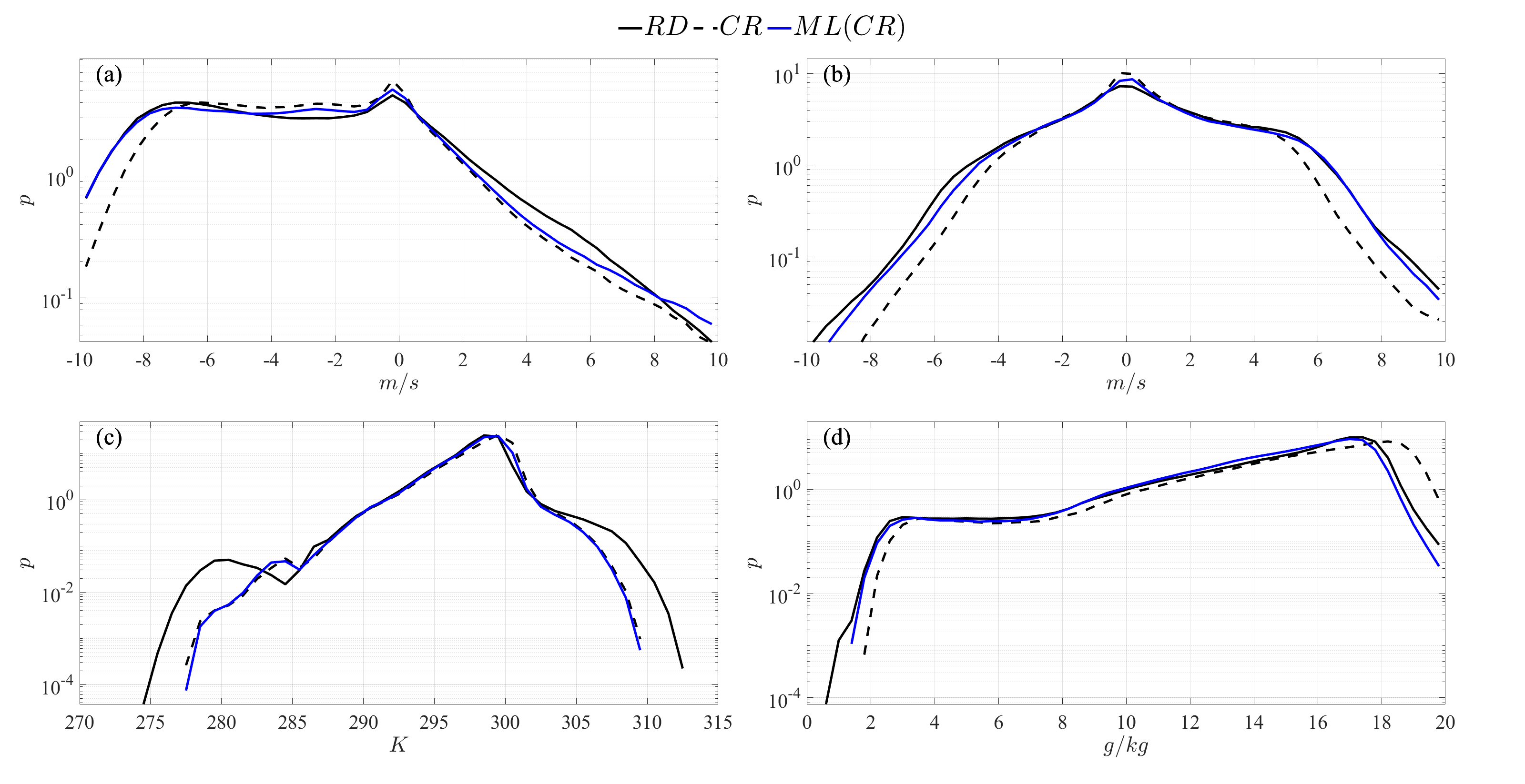}
    \caption{36 year probability density function for surface sigma-level 72 in the tropics.  $U$ (a), $V$ (b), $T$ (c), $Q$ (d). Results are shown for ERA5 reanalysis data (RD), free-running data (CR), and ML corrections.}
    \label{fig:Tropics_72}
\end{figure}

\begin{figure}
    \centering
    \includegraphics[ width=1\textwidth]{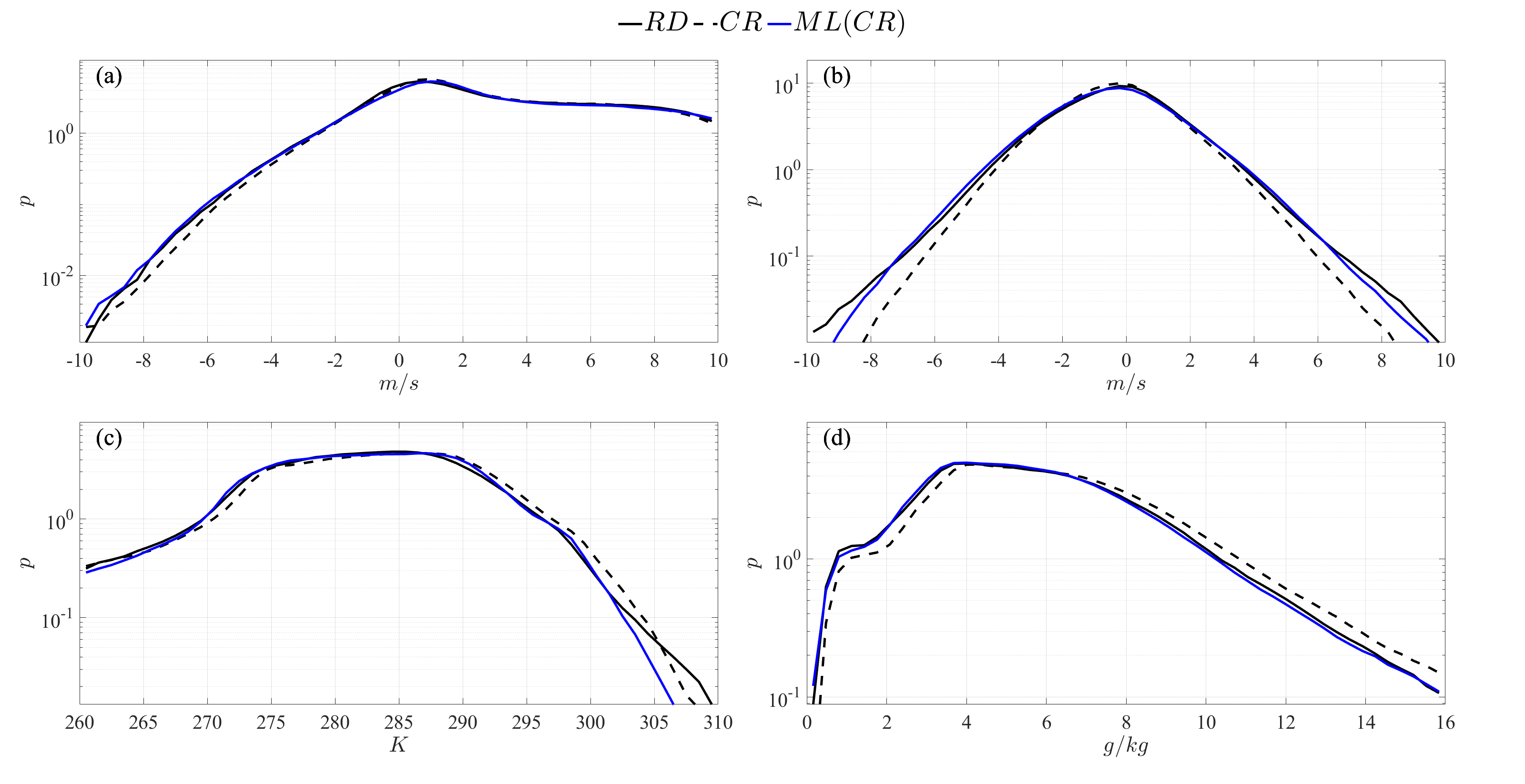}
    \caption{36 year probability density function for surface sigma-level 72 in the mid-latitude region.  $U$ (a), $V$ (b), $T$ (c), $Q$ (d). Results are shown for ERA5 reanalysis data (RD) (solid black), free-running data (CR) (dashed black), and ML corrections (blue).}
    \label{fig:MidLat_72}
\end{figure}

\begin{figure}
    \centering
    \includegraphics[ width=1\textwidth]{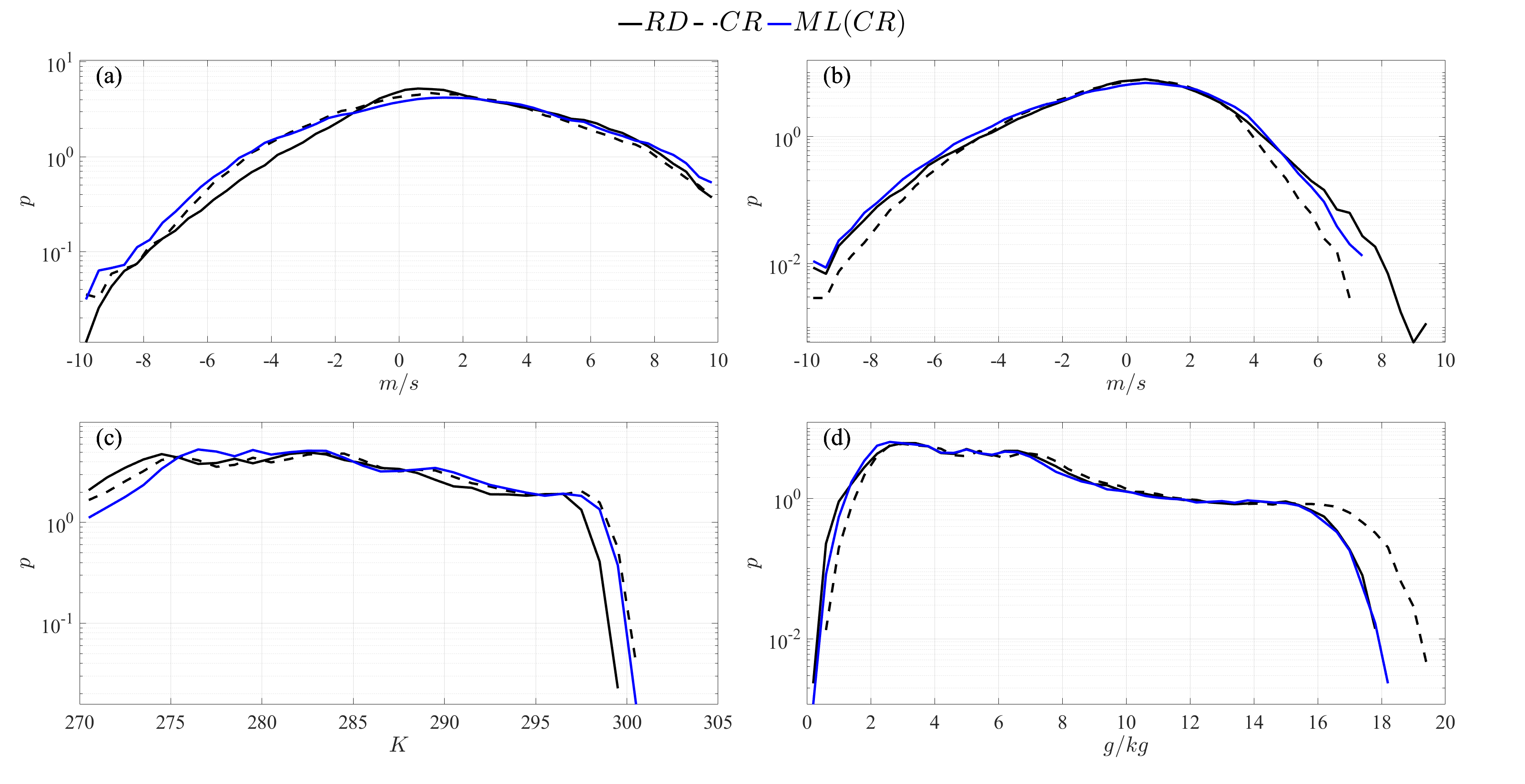}
    \caption{36 year probability density function for surface sigma-level 72 in northwest Pacific.  $U$ (a), $V$ (b), $T$ (c), $Q$ (d). Results are shown for ERA5 reanalysis data (RD) (solid black), free-running data (CR) (dashed black), and ML corrections (blue).}
    \label{fig:NWP_72}
\end{figure}

In addition to global statistics, policy makers preparing for the increased risks of climate change require accurate risk analysis over a range of spatial scales. Therefore we also analyze the statistics of the predicted climate over several regions of varying size: the tropics, mid-latitude, continental US, northeast US, northern Europe, and the northwest Pacific. The size and location used in the following results are summarized in table \ref{tab:regions}. As in \S\ref{sec:global_stats} we focus on sigma level 72, the level closest to the surface. Figures \ref{fig:Tropics_72} - \ref{fig:NWP_72} show the probability density functions of the four progress variables $U,V,T$, and $Q$ in the tropics, mid-latitude, and the northwest Pacific regions. Result for the remaining regions are included in appendix \ref{app:additonal_E3SM_stats}. The reanalysis reference is shown in solid black, the free-running E3SM and ML correction thereof are shown in dashed black and blue respectively. Again, we see that the ML correction is most pronounced in regions where the E3SM model alone is most biased. Most notably the specific humidity $Q$ (subplot d in figures \ref{fig:Tropics_72} - \ref{fig:NWP_72}) and meridional wind speed $(V)$ (subplot b in figures \ref{fig:Tropics_72} - \ref{fig:NWP_72}) where for all regions the ML correction brings the tails of the predicted distribution into good agreement with ERA5 data. See also figure \ref{fig:Tropics_72}a, where the ML correction does significantly improves the prediction of the zonal wind speed $(U)$. As with the global statistics, the ML correction has only minor impacts on the distributions of temperature $(T)$. However, with the exception of the tropics region (figure \ref{fig:Tropics_72}c) this is generally well predicted by the E3SM model alone and notably in no region does our ML correction significantly increase bias. The fact that our correction operator is able to improve predictions across all variables and over a range of spatial scales is a promising result, as it shows that the predicted flow field could in principle be further used for targeted super-resolution to predict local features on scales smaller than than the grid of the coarse model.

\begin{table}[]
    \centering
    \begin{tabular}{c|c|c}
     \textbf{Region}  & \textbf{Latitude} & \textbf{Longitude}  \\
     \hline
     Mid-latitude   & $30S-60S$ $\&$ $30N-60N$ & $0-360$ \\
     \hline
     Tropics  & $20S-20N$ & $0-360$ \\
     \hline
     Continental US   & $25N-55N$ & $90W-120W$ \\
     \hline
     Northeastern US   & $25N-55N$ & $60W-90W$ \\
     \hline
     Northern Europe   & $40N-70N$ & $10E-40E$ \\
     \hline
     Northwest Pacific   & $30N-60N$ & $150E-180E$ \\
     \hline
    \end{tabular}
    \caption{Summary of regions analyzed in \S\ref{sec:regional_stats}}
    \label{tab:regions}
\end{table}

\section{Discussion}\label{sec:discussion}
We have introduced a method to machine learn correction operators to improve the statistics of under-resolved simulations of turbulent dynamical systems. The premise of the proposed strategy is to generate training data pairs which are minimally affected by chaotic divergence. Instead of using an \textit{arbitrary} coarse trajectory as the training input, we used a coarse trajectory \textit{nudged} towards the training target trajectory. This nudged trajectory predominately obeys the dynamics of the coarse model, yet is constrained from \textit{randomly wandering} too far from the reference. In essence, it is an approximation of the one (of infinitely many) trajectory of the coarse model which is closest to the reference data. Once trained on this specific pair of trajectories, an ML operator can reliably map \textit{any} free-running coarse trajectory into the attractor of the reference data. The critical benefit of such an operator is that it acts on data in a post-processing manner, and is thus unaffected by the stability issues, and practical implementation challenges, which plague machine learned corrections of the system dynamics.

A key aspect of the proposed approach is the ability to incorporate, directly into the learning process, dynamical information that goes beyond statistics of the training data. This is achieved through an objective function that is matching \textit{trajectories} rather than their statistics. This is critical especially for extreme events, where the key information `lives' in the very structure of the trajectory over the short duration of such events. Cost functions formulated to match statistics, either need to incorporate high order statistical information (something that is practically impossible because of both inadequate data but also vast computational cost) or they are doomed to have poor generalization properties since low order statistics (e.g. spectrum) cannot `see' the dynamics of extreme events. On the other hand, the formulated approach eliminates the divergence due to chaotic behavior and uses the maximum information from the reference data by training \textit{in the time domain}, i.e. directly fixing the structure of the trajectory near an extreme event. This allows for unprecedented improvement especially for extreme event statistics. 

The proposed strategy was first illustrated on a prototypical two layer quasi-geostrophic climate model using a simple LSTM network architecture. In this reduced order system our ML correction operator was able to bring the global, and scale-by-scale statistics of a severely under-resolved simulation, simulated on a $24\times24$ grid, into good agreement with the fully-resolved reference solved on a $128\times128$ grid. Additionally, we demonstrated the ability to accurately predict statistics for time horizons much longer than the training data, and for parameter regimes outside of that training data. We then applied our framework to a realistic climate model -- the Energy Exascale Earth System Model (E3SM) solved on a grid with approximately $110$ km horizontal resolution. In this case the reference data used as the training target and the evaluation metric was not a fully resolved simulation, but ERA5 reanalysis data. To address this far more complex system, we designed a network architecture which combined the LSTM base we used for the simpler QG system with overlapping convolutional layers used to extract local anisotropic features from the input data. We found that our ML correction significantly reduced the bias of the E3SM solution, bringing the statistics of \textcolor{black}{the wind speeds and specific humidity into good agreement with reanalysis data on both a global and regional level. The debiasing capabilities of our ML correction were less pronounced in the case of temperature, for which the improvements, especially in the tails of the distributions were more modest, and more region dependent.  The improvement in the wind speed and humidity statistics however are especially notable as these } variables were not well approximated by the free-running E3SM solution. In particular, the correction operator significantly improved the predictions of the tails of \textcolor{black}{these} distributions which are critical for quantifying the risks of extreme weather events. In addition to the primitive variables, we also analyzed the mean integrated vapor transport (IVT), a highly anisotropic integral quantity of particular practical interest as it drives atmospheric rivers and thus precipitation. Here the improved predictions in the wind speed and humidity of our ML correction combined to reduce the overall RMSE in IVT by 51$\%$, and successfully removed several systematic regional biases of the coarse model, such as its tendency to underpredict the vapor transport in the southern hemisphere.

While the proposed methodology was demonstrated to be effective for the prediction of a multitude of climate metrics, some limitations of the current setup should be stated. First, the approach works well under the assumption that the climate is in a statistically steady \textcolor{black}{state}, for which a mapping can be learned through the proposed training scheme. Hence, applying the learned model in situations where the climate undergoes a transitory phase may hinder its performance, unless similar transitory intervals are included in the training data. This is particularly true if the transition is not captured at all by the coarse-scale model. Furthermore, when applied to future climate scenarios with drastically different forcing, the requirement for reference data -- which may not be available at high resolution for long times -- makes it difficult to assess the predictive powers of our approach a priori. For such runs to be included in training, high-fidelity simulations would have to be used as reference and the coarse models nudged towards them. This limitation however is true for online data-driven correction schemes as well since most such models lack concrete error bounds for out-of-sample predictions. Furthermore, for the application of the scheme to dynamical systems broadly, there is no guarantee that a nudged simulation exists that follows the reference data closely while \textcolor{black}{satisfying the dynamics} of the coarse simulation. Essentially, if the coarse model is too far from the reference data, i.e. too under-resolved or neglecting too much important physics there is no guarantee the process will work.

One of the main advantages of the proposed framework is its generality and non-intrusive nature. Theoretically, intrusive online approaches act on the dynamics of the system, but practically, this means they act on \textit{software}, i.e. they must be integrated with existing code stacks.  For modern ESMs, this code stack can be complex or proprietary, making the implementation of such strategies difficult or even impossible if the source code is unavailable. On the other hand, non-intrusive approaches, such as the one proposed here, act on \textit{data} -- meaning the model is agnostic to the specific software implementation of the model generating the data. Generating the training data does require implementing a nudging tendency in the climate model code, however, this is generally a much less invasive task than integrating an ML operator, which may be implemented in a different software language than the climate model itself \cite{mcgibbon_fv3gfs-wrapper_2021}. Then once trained the model can be used without further intrusion into the core ESM. Another strength, is that the proposed framework provides predictions of all progress variables, $(U,V,T,Q)$, at all grid points and all sigma levels -- a feature not shared by all debiasing schemes. This in turn means that the flow fields predicted by our correction operator could then be used for local super-resolution (down-scaling) to investigate local climate forecasting and impact assessment. However, further work is required to \textcolor{black}{investigate the ability of our approach to improve the statistics of other climate metrics such as precipitation and to} ensure that the corrected fields obey basic physical constraints such as geostrophic balance or conservation of mass and energy over the spatio-temporal scales relevant to such local analysis. We believe that by lowering these barriers to adoption, our approach has the potential to significantly accelerate and democratize the implementation of data-driven climate modeling. To this end, extensions of our approach such as built in uncertainty quantification, physics informed constraints, and grid-agnostic network architectures -- which could allow for applications across different ESMs -- are the topic of ongoing research.

\section{Acknowledgments}
This research has been supported by the DARPA grant HR00112290029 under the program `AI-Assisted Climate Tipping Point Modeling' supported by the Program Manager Dr. Joshua Elliott. Pacific Northwest National Laboratory is operated for the U.S. Department of Energy by Battelle Memorial Institute under Contract DE-AC05-76RL01830. Computational resources for the material shown in this work were provided by Anvil super computer through the ACCESS program. The authors thank Prof. G. Karniadakis for stimulating discussions on this work. We are also grateful to Dr. S. Khurshid for advise and support on using the Anvil super computer.

\section*{Open Research Section}

The source code for the E3SM \cite{eam_code} climate model used to generate the simulations discussed in \S\ref{sec:E3SM} was obtained from the Energy Exascale Earth System Model project, sponsored by the U.S.Department of Energy, Office of Science, Office of Biological and Environmental Research. The ERA5 reanalysis data used as a reference for training the ML model and generating the reference data in \S\ref{sec:E3SM} is available at the Copernicus Climate Change Service (C3S) Climate Data Store via https://doi.org/10.24381/cds.bd0915c6  \cite{Hersbach_et_al:2020}.
The software and data needed to generate the results described here can be found on Zenodo at  \url{https://zenodo.org/doi/10.5281/zenodo.10657047} \cite{barthel_data_2023}.

\FloatBarrier
\appendix
\section{Appendix}

\subsection{Nudging Implementation in E3SM}\label{app:nudg_workflow}
Here we briefly outline the practical implementation of the nudging strategy in the E3SM model used to train the ML correction operator used to generate the results in \S\ref{sec:E3SM}. We follow the formulation of \citeA{sun_impact_2019} and \citeA{zhang_further_2022}, for which the nudged governing equations of the E3SM model takes the form
\begin{eqnarray} \label{eqn:ndg_forcing}
\frac{\partial \boldsymbol{X}}{\partial t} = 
       \underbrace {\boldsymbol{D} \left(\boldsymbol{X} \right)}_{dynamics} 
     +  \underbrace {\boldsymbol{P} \left(\boldsymbol{X} \right)}_{physics} 
     - \underbrace {\boldsymbol{\mathcal{N}}\left(\boldsymbol{X},\boldsymbol{X}^{RD}\right)}_{nudging} 
\end{eqnarray}
where $\boldsymbol{D}$ represents the resolved dynamics, $\boldsymbol{P}$ represents the parameterized physics and $\boldsymbol{\mathcal{N}}$ is the nudging tendency. The nudging tendency is applied at each grid point and is specifically implemented as
\begin{equation}
  \label{eqn:ndg_term}
      \boldsymbol{\mathcal{N}}\left(\boldsymbol{X},\boldsymbol{X}^{RD}\right) = 
       \begin{cases}
         0                                                                                , & \text{if } P  \le 1~$\rm Pa$ \\
         \\
       - \dfrac{ \boldsymbol{X} - \boldsymbol{X}^{RD}}{\tau} \times \dfrac{P_m}{P_{0}}, & \text{if } 1~$\rm Pa$ < P \le P_{0} \\
         \\
       - \dfrac{ \boldsymbol{X} - \boldsymbol{X}^{RD}}{\tau} \times \dfrac{1}{2}\left[ 1 + \mathrm{tanh} \left( \dfrac{Z- Z_b}{0.1Z_b}\right)\right] , & \text{if } Z  \le Z_p \\ 
         \\ 
       -  \dfrac{ \boldsymbol{X} - \boldsymbol{X}^{RD}}{\tau} , & \text{otherwise} 
    \end{cases}
\end{equation}
where $\boldsymbol{X} = (U,V,T,Q)$ is the state variable, $\boldsymbol{X}^{RD}$ is the ERA5 reference, $P_m$ and $Z_m$ represent the atmospheric pressure and geopotential height at a given sigma level, and $\tau$ denotes the relaxation time scale. Following \citeA{sun_impact_2019} and \citeA{zhang_further_2022} we fix $\tau = 6$hr. The simulation uses a time step of 0.5hr and the ERA5 reference data is defined at 3-hourly increments and interpolated at each time step using the linear temporal interpolation described in \citeA{sun_impact_2019}. 
The quantities $P_0$ and $Z_b$ are user defined threshold parameters which govern how the nudging tendency is modulated in the upper and lower ends of the atmosphere. $Z_b$ is set at the  planetary boundary layer height (PBLH), which is diagnosed and dynamically set at each time step. $P_0$ is set to $30$Pa, $30$Pa, $10$Pa, and $100$Pa for the variables $U,V,T,Q$ respectively and held constant throughout the simulation. This modulation in the upper and lower sigma levels differs from the default formulation proposed by \citeA{sun_impact_2019} and \citeA{zhang_further_2022}, however, it is implemented here to account for uncertainties in our specific reference data. We de-emphasize the nudging tendency in the upper atmosphere due to the deteriorating quality of the ERA5 reanalysis data at those altitudes, while near-surface the concern is the significant errors which arise over the high-terrain regions when ERA5 data is mapped onto the E3SM model grid.

\subsection{Additional E3SM Results}\label{app:additonal_E3SM_stats}
\begin{figure}
    \centering
    \includegraphics[width=1\textwidth]{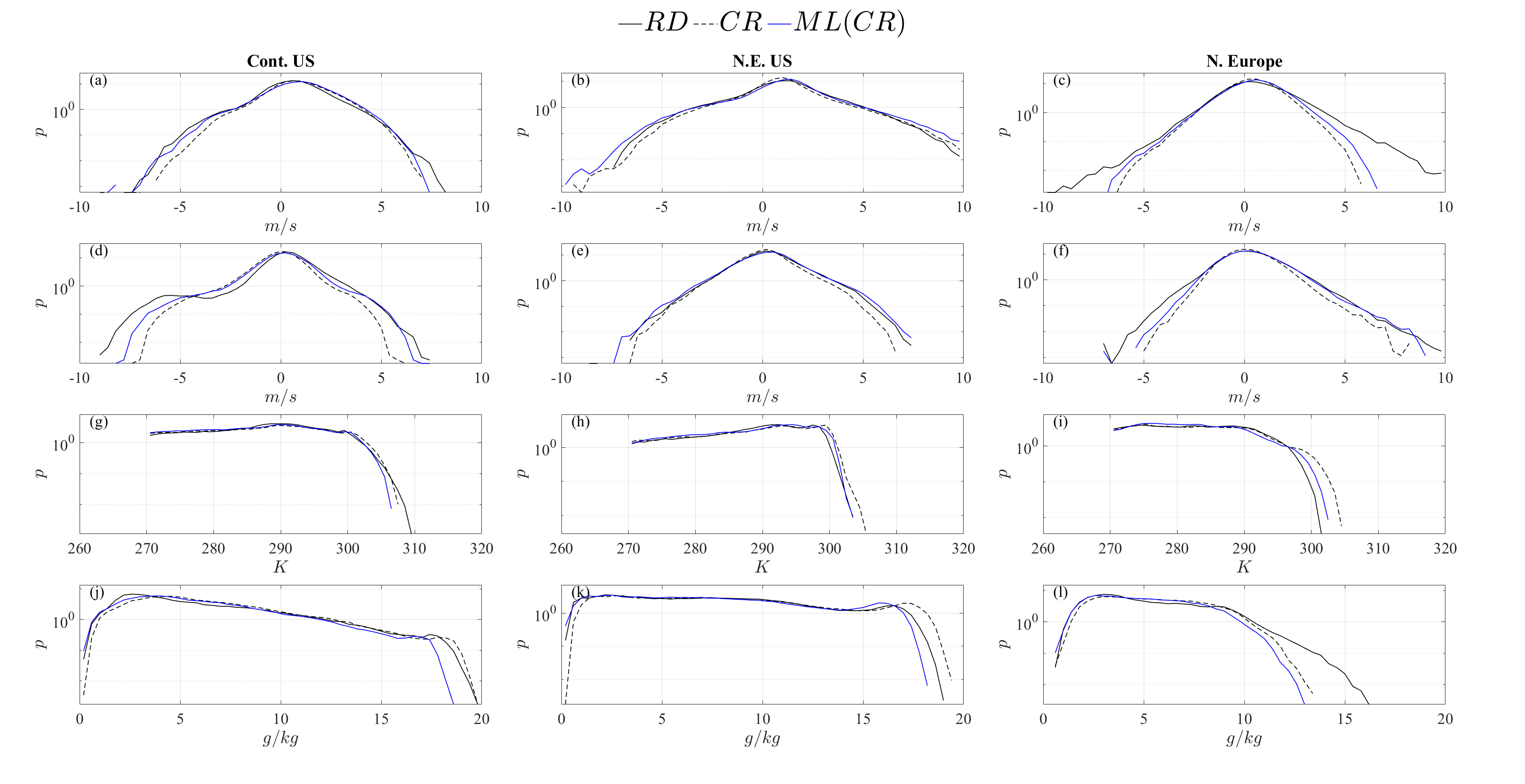}
    \caption{30 year probability density function for surface sigma-level 72 for Continental US (left column), northeastern US (center column) and northern Europe (right column).  $U$ (a,b,c) and $V$ (d,e,f), $T$ (g,h,i), $Q$ (j,k,l). Results are shown for ERA5 reanalysis data (RD) (solid black), free-running data (CR) (dashed black), and ML corrections (blue).}
    \label{fig:pdf_appendix_72}
\end{figure}

Here we show some additional results for \S\ref{sec:E3SM}. Figure \ref{fig:pdf_appendix_72} shows the regional probability density functions for the regions not shown in \S\ref{sec:E3SM}: Continental US (left column), northeastern US (center column) and northern Europe (right column) at the surface  sigma level 72. 

\FloatBarrier
\newpage

\bibliography{references_no_url,references_alexis}
\end{document}